\documentclass[12pt,aps,english]{revtex4}
\usepackage{graphicx}
\usepackage{bm}
\usepackage{dcolumn}
\usepackage[usenames, dvipsnames]{color}
\usepackage{epsfig}
\usepackage{amsmath}
\usepackage{mathrsfs}

\begin{document}

\title{Transport of probe particles in polymer network: effects of probe size, network rigidity and probe-polymer interaction}
\author{Praveen Kumar, Ligesh Theeyancheri, Subhasish Chaki and Rajarshi Chakrabarti*}
\affiliation{Department of Chemistry, Indian Institute of Technology Bombay, Mumbai, Powai 400076, E-mail: rajarshi@chem.iitb.ac.in}
\date{\today}

\begin{abstract}
\noindent  
Fundamental understanding of the effect of microscopic parameters on the dynamics of probe particles in different complex environments has wide implications. Examples include diffusion of proteins in the biological hydrogels, porous media, polymer matrix, etc. Here, we use extensive molecular dynamics simulations to investigate the dynamics of the probe particle in a polymer network on a diamond lattice which provides substantial crowding to mimic the cellular environment. Our simulations show that the dynamics of the probe increasingly becomes restricted, non-Gaussian and subdiffusive on increasing the network rigidity, binding affinity  and the probe size. In addition, the velocity autocorrelation functions show negative dips owing to the viscoelasticity and caging due to surrounding  network.  These observations go with general experimental findings. Surprisingly for a probe particle of size comparable to the mesh size, unrestricted motion engulfing large length scales has been observed. This happens with the more flexible polymer network, which is easily pushed by the bigger probe. Our study gives a  general qualitative picture of transport of probes in gel like medium, as encountered in different contexts.
\end{abstract}

\maketitle

\section{Introduction}
\noindent The dynamics of particles in complex fluids is a subject of fundamental interest in physics \cite{barkai2012single}, chemistry \cite{devetter2014observation}, materials science  \cite{skaug2015hindered} and biology \cite{majumdar2019liquid, di2014probing, werner2019drag,chakrabarti2014diffusion}. This is an omnipresent situation, where a tracer particle transports through a crowded environment and its mean squared displacement (MSD) scales as $t^\alpha$ \cite{ghosh2016anomalous, samanta2016tracer, kalathi2014nanoparticle, sprakel2008brownian, chatterjee2011subdiffusion, Chakrabarti2013} with $\alpha < 1$ so that the dynamics is subdiffusive. A purely diffusive regime is when $\alpha=1$ and generally achieved in the long time limit or in the absence of any crowders. \textit{In-vivo} examples involve tracer particles in the cytoplasmic fluid of living cells \cite{norregaard2017manipulation}, intracellular transport of insulin granules \cite{tabei2013intracellular}, anomalous diffusion of telomeres in the nucleus of mammalian cells \cite{bronstein2009transient}, diffusion of protein through nuclear pore complex (NPC) \cite{chatterjee2011subdiffusion, goodrich2018enhanced, Chakrabarti2013, gopinathan2017}  and mucus membrane \cite{lieleg2012mucin} etc. Other \textit{in-vitro} examples include nanoprobe motion in polymer matrix \cite{nahali2018nanoprobe}, gel \cite{godec2014collective, seiffert2010controlled}, polymer films \cite{flier2011heterogeneous, bhattacharya2013plasticization}.  For another class of systems where the dynamics are driven by some external or internal energy consumption into directed motion \cite{du2019study}, $\alpha$ has been found to be greater than 1 and termed as superdiffusion \cite{chaki2019effects}. For example, polymer chain in active bath \cite{chaki2019enhanced, samanta2016chain}, particle diffusion in bacterial bath \cite{wu2000particle}, energy consuming catalytic enzymes \cite{jee2018catalytic, mohajerani2018theory}. 
\\
\\
Experimental techniques such as Fluorescence Correlation Spectroscopy (FCS) \cite{hofling2011anomalous, nandy2019structure}  and single particle tracking \cite{bhattacharya2013plasticization,woll2009polymers} are routinely used to monitor the spatio-temporal dynamics of small molecules and nanoprobes in polymeric environments or biological cells. Such experiments have shown that the tracer dynamics in such complex medium is non-Gaussian in addition to being subdiffusive \cite{zhou2008macromolecular,Xue2016,bhowmik2018non}. The non-Gaussianity and subdiffusive behavior are often short time phenomena and arises due to heterogeneity in the medium. For example, one often observes anomalous diffusion of artificially introduced tracer particles in the cytoplasmic fluid of living cells \cite{norregaard2017manipulation}.  Another important, yet less explored, aspect is the nature of the anomalous diffusion in bio-gel, which are polymer networks consisting of actin and other biofilaments through which biomolecules diffuse \cite{phillips2012physical, mizuno2007nonequilibrium, sonn2017scale}. Bio-polymer gels are ubiquitous in living organisms. Except for bones, teeth and nails, mammalian tissues are largely gel-like materials that are mainly composed of protein and polysaccharide networks with a water content of up to 90\% \cite{cherstvy2019non}. The morphology of the matrix, i.e, the mesh-like structure and sticky interaction of polymer allows the passage of certain molecules like signaling proteins, nutrients and drugs \cite{lai2009mucus} and can reject bacteria, toxic agents etc \cite{mcguckin2011mucin}. Thus, sticky polymer-based mucus hydrogels are robust and serve as selectively-permeable biological particle filters that play a crucial role in tissue protection \cite{lieleg2012mucin} and cell functioning of human and animal bodies \cite{thornton2008structure}. However, for attractive (or sticky) interaction, the tracer particle can bind with the polymer which eventually slows down the diffusion process \cite{hansing2018particle,carroll2018diffusion}. At long time, the polymer matrix relaxes and the tracer follows free diffusion.  Alternatively, for repulsive (or nonsticky) interaction between tracer and polymer gel, diffusion will be much faster than free Brownian motion which enables the organism to effectively transport biomolecules \cite{tuteja2007breakdown,grabowski2009dynamics}.  However, recently a mechanism has been proposed, where the transporting particles bind to a crosslinking of a polymer gel and breaks the crosslinking in the long time scale and shows enhanced diffusion \cite{goodrich2018enhanced}. This goes hand-in-hand with the earlier theoretical prediction of faster diffusion of proteins through gel like central plug of NPC, facilitated by the fluctuations of the gel \cite{chakrabarti2014diffusion}.
\\
\\
The density of the polymer network \cite{johansson1991diffusion}, chain stiffness \cite{TaeJung2011}, solute size \cite{goodrich2018enhanced} and the geometrical arrangement of the polymer chains \cite{godec2014collective} are  major factors in the regulation of numerous cellular processes \cite{netz1997computer}. Recently, it has been observed that crowding affects protein folding and stabilization \cite{ping2003effects}, gene expression \cite{norred2018macromolecular}, cellular signaling \cite{hellmann2012enhancing}, and conformational transition of macromolecules \cite{samiotakis2009folding}. Inside a cell, molecular crowding may reach a volume occupation of up to 40 \% and thus lead to the slowing-down of diffusion \cite{norregaard2017manipulation, konopka2006crowding, barkai2012single}. Indeed, it has been observed that particles inside the living cell exhibit anomalous diffusion with the scaling exponent $\alpha$ in the subdiffusive range \cite{barkai2012single}.  A wide range of $\alpha \sim 0.4-0.9$ has been reported for the motion of membrane protein and lipids \cite{horton2010development}, messenger RNA molecules in E. coli bacteria \cite{golding2006physical}, lipid granules \cite{jeon2011vivo}, chromosomal loci \cite{weber2010bacterial}, hair bundles in ears \cite{kozlov2012anomalous} etc. While crowding would be expected to hinder the particle's mobility \cite{dix2008crowding}, it enhances the search process of reactive proteins for colliding with each other, essentially increasing the rate of biochemical reactions \cite{minton1992confinement}.  In addition, crowding can change the free volume of the polymer gel in response to external stimuli such as a change in temperature, humidity, and  pH \cite{sahoo1998ph,park1999temperature,bhattacharya2013plasticization}. 
\\
\\
\noindent Often the dynamics of tracer particles in polymeric environment (solution or gel) is different on short and long time scales \cite{chen2019influence}. At relatively short time scales, the polymer matrix imposes barriers to tracer's diffusion within the void space \cite{dell2014theory} and the motion of the tracer turns out to be transiently subdiffusive \cite{hofling2013anomalous}. On the long time scale, structural reorganization happens within the gel and the MSD crosses over to Brownian motion  \cite{piskorz2014universal, sokolov2012models, orlandini2019polymerization}. Moreover, it is experimentally observed that the distribution of the tracer's displacement is not always Gaussian for Brownian diffusion \cite{wang2012brownian,wang2009anomalous}. The observed anomaly in such systems is addressed by various stochastic processes. These include continuous time random walks (CTRW) and fractional Brownian motion (FBM) \cite{bouchaud1990anomalous,scher1975anomalous,deng2009ergodic,goychuk2009viscoelastic}. CTRW models are closely related to temporary cages formed by the polymers (or the crowders) whereas FBM is typically associated with the motion of a random walker in a viscoelastic medium \cite{klafter2011first}. However, the physical origin of the non-Gaussianity in the displacement distribution remains an open question \cite{metzler2014anomalous}. This has been rationalized with the hypothesis that a tracer can have a distribution of random diffusivities which can lead to a slower (or caged) and faster motion in a complex environment \cite{chubynsky2014diffusing,jain2016diffusion,Kwon2014,chechkin2017brownian,acharya2017fickian,Lanoiselee2018}. Still, a consensus is lacking on the physical picture of the anomalous diffusion and non-Gaussian distribution of passive tracer particles in a complex and crowded environment \cite{barkai2012single}.  \\

\noindent In this paper we focus on elucidating the effects of probe or tracer size, probe-polymer interaction and network stiffness on the nanoprobe transport through a polymer-network. Biological cells, membranes provide a gel-like environment and nanoprobes are often biomolecules such as proteins, RNA \cite{norregaard2017manipulation, lieleg2012mucin, goodrich2018enhanced, godec2014collective, seiffert2010controlled, nandy2019structure}.  We look at the transport of a Lennard-Jones probe in a polymer network on a diamond lattice, which provides substantial crowding. In particular, we emphasize on how the affinity of the nanoprobe to the network influences the transport. Specifically, increasing the stickiness or the binding affinity of the probe to the network leads to caging resulting confined non-Gaussian subdiffusion in the short to moderate time. On the other hand, on increasing the stiffness of the network as one may encounter in polymer hydrogel with low humidity content, we see narrower displacement distribution as observed in single molecule tracking experiments \cite{bhattacharya2013plasticization}.  Our observation for large probes comparable to network mesh size is quite interesting. Our simulations show that moderately sticky larger probes in a relatively flexible polymer network stretches the network and  has a finite but small probability to make large amplitude motion. Smaller tracers do not show such mode of transport.  On increasing the stiffness of the network, even bigger probes cannot efficiently stretch the network and the large displacement motion ceases. \\

\noindent The paper is arranged as follows. In Section \ref{sec:simulation} we present the simulation details. The results and discussion are given in Section \ref{sec:results} and we conclude the paper in Section \ref{sec:conclusion}. 

\section{Simulation details} \label{sec:simulation}

\begin{figure}
	\includegraphics[width=0.8\textwidth]{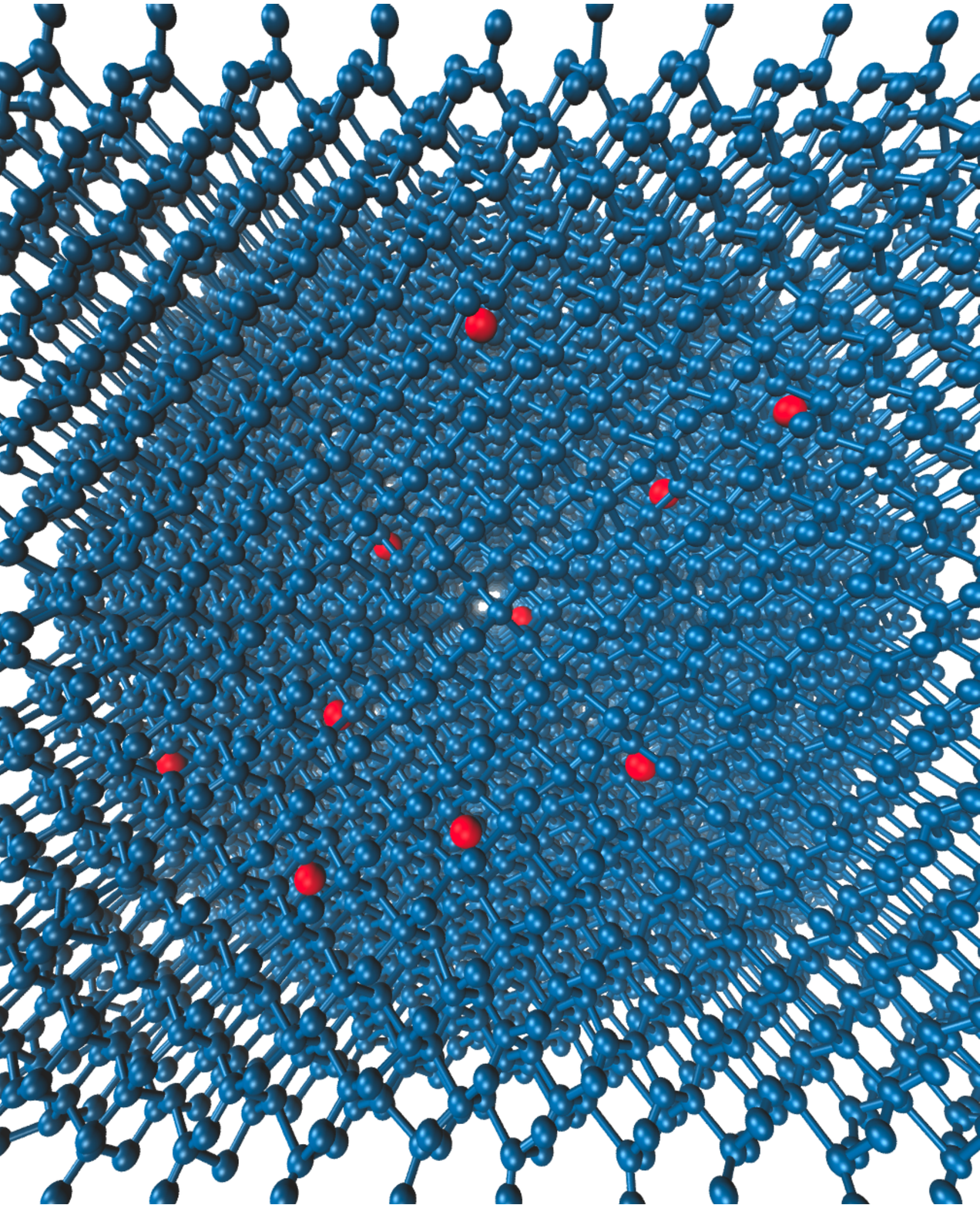}
	\caption{A snapshot of the tracers (red) inside the polymer network (blue). The snapshot is created with Visual Molecular Dynamics (VMD) package \cite{humphrey1996vmd}. Only a part of the polymer network is shown. }
	\label{fig:gel}
	\end{figure}

\noindent The simulations are carried out using LAMMPS \cite{plimpton1995fast}, a freely available open-source molecular dynamics package. In describing the model system, the Lennard-Jones parameters $(\sigma$ and $\epsilon)$ and mass $m$ are the fundamental units of length, energy and mass respectively. All the particles in the system have identical masses $(m)$. Accordingly, the unit of time is $\tau=\sqrt{\frac{m \sigma^2}{\epsilon}}$. All other physical quantities are therefore reduced accordingly, expressed in terms of  these fundamental units, $\sigma, \epsilon$ and $m$ and presented in dimensionless forms. The polymer network is created on a diamond lattice and consists of $9883$ number of monomers, each of diameter $\sigma$. The lattice coordinates are generated using an open-source package VESTA (Visualization for Electrical and Structural Analysis) \cite{momma2011vesta}. Thus each lattice site has a monomer and each monomer is connected to four neighboring monomers through finitely extensible nonlinear elastic (FENE) spring. A snapshot of the gel is presented in Fig. (\ref{fig:gel}). The FENE potential is as follows.

\begin{equation}
V_{FENE}\left(r_{ij} \right)=\begin{cases} -\frac{k r_{max}^2}{2} \log\left[1-\left( {\frac{r_{ij}}{r_{max}}}\right) ^2 \right],\hspace{5mm} \mbox{if } r_{ij} \leq r_{max}\\
=\infty, \hspace{41mm} \mbox{otherwise}
\end{cases}
\label{eq:FENE}
\end{equation}
where $r_{ij}$ is the distance between two neighboring monomers in the polymer network with a maximum length $r_{max} = 2.5 \sigma$ and $k$ is the force or the stiffness constant, a measure of the network stiffness and has the unit of $\frac{\epsilon}{\sigma^2}.$ 
\\
\\
The non-bonded attractive interactions between monomers of the polymer network and tracers are modeled by Lennard Jones (LJ) potential:
\begin{equation}
V_{LJ}^{ij}(r_{ij})=\begin{cases}4\epsilon_{ij}\left[\left(\frac{\sigma_{ij}}{r_{ij}}\right)^{12}-\left(\frac{\sigma_{ij}}{r_{ij}}\right)^{6}\right], \hspace{5mm} \mbox{if } r_{ij} \leq r_{cut}^{ij}\\
=0, \hspace{41mm} \mbox{otherwise}\\
\end{cases}
\label{eq:LJ}
\end{equation}	
In the above expression, the subscripts $i$ and $j$ represent both the monomers and the tracers, $r_{ij}$ is the separation between two particles $i$ and $j$ and $\epsilon_{ij}$ is strength of the attractive interaction or binding affinity, $\sigma_{ii}$ is the diameter of the particle $i$ and $\sigma_{ij}$ is the sum of the radii of two interacting particles, $\sigma_{ij} = \frac{1}{2}(\sigma_{i}+\sigma_{j})$ and $r_{cut}^{ij} = 2.5 \sigma_{ij}$ is the cutoff radius for monomer-tracer pair interaction.
\\
\\
The monomer-monomer and the tracer-tracer interactions are set as purely repulsive and modeled by the Weeks–Chandler–Andersen (WCA) potential \cite{weeks1971role}:
\begin{equation}
V_{WCA}(r_{ij})=\begin{cases}4\epsilon_{ij}\left[\left(\frac{\sigma_{ij}}{r_{ij}}\right)^{12}-\left(\frac{\sigma_{ij}}{r_{ij}}\right)^{6}\right]+\epsilon_{ij}, \hspace{5mm} \mbox{if } r_{ij} \leq (2)^{1/6}\sigma_{ij}\\
=0, \hspace{50mm} \mbox{otherwise}
\end{cases}
\label{eq:WCA}
\end{equation}
For WCA $i=j$ and $\epsilon_{ii}=\epsilon.$
\\
\\
For each simulation, the system consists of 10 tracers, which are packed into a cubic box of length $38 \sigma$. Periodic boundary conditions are set in all the three directions. The time step $\delta t$ = 0.001$\tau$ is chosen to be a constant in all the simulations. After equilibrating the system long enough so that the average monomer-monomer distance is nearly constant and found around $1.12\sigma$ (not shown), which is also a crude measure of the mesh size for the polymer network. All the production simulations are carried out for 3 $\times 10^6$ steps.  The positions and velocities of the tracer particles are saved every $10$ steps. All the simulations are performed using the Langevin thermostat and equation of motion integrated using velocity Verlet algorithm in each time step. 
\\
\\
We implement following underdamped Langevin equation to simulate the motion of  the $i^{th}$ particle of our system with mass $m$ with the position $r_{i}(t) $ at time t:
	\begin{equation}
	m\frac{d^2 r_{i}(t)}{dt^2}=-\xi\frac{dr_{i}}{dt}-\bigtriangledown\sum_{i} V(r-r_{i})+f_{i}(t)
	\label{eq:langevineq}
	\end{equation} 
Where $\xi$ is the friction coefficient and $\xi=1$, $m=1$ in all the simulations, $f_{i}(t)$ is the Gaussian thermal noise with the statistical properties,
	\begin{equation}
	\left<f(t)\right>=0, \hspace{5mm}
	\left<f_{\alpha}(t^{\prime})f_{\beta}(t^{\prime\prime})\right>=6 \xi k_B T \delta_{\alpha\beta}\delta(t^{\prime}-t^{\prime\prime})
	\label{eq:random-forcerouse}
	\end{equation}
where $k_{B}$ is the Boltzmann constant, T is the temperature and $ \delta$ represents the Dirac delta-function, $\alpha$ and $\beta$ represent the cartesian components. We consider the thermal energy $k_B T=1$. In our simulations, we choose four different values of $\epsilon_{ij} (=2, 3, 4, 5)$, four different values of tracer sizes $\sigma_{ii} (=0.3, 0.5, 1, 1.5 )$ and three different values of $k (=5, 10, 15)$. We do not include the hydrodynamic interaction in our simulations.

\section{Results and discussion} \label{sec:results}

\subsection{Mean square displacement, time exponent and long-time diffusivity}

\begin{figure}[h]
	\centering
	\begin{tabular}{ccc}
		\includegraphics[width=0.33\textwidth]{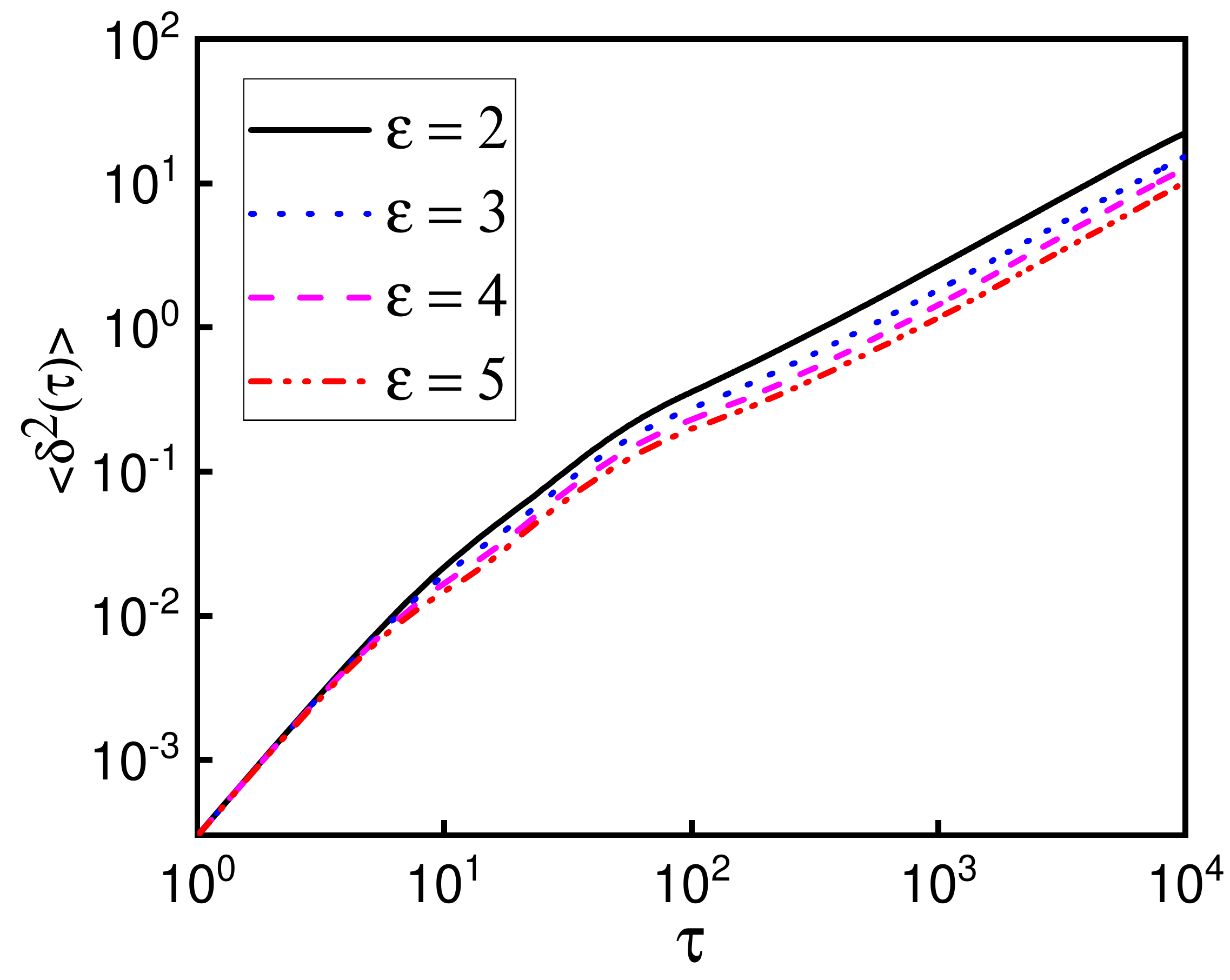} & 
		\includegraphics[width=0.33\textwidth]{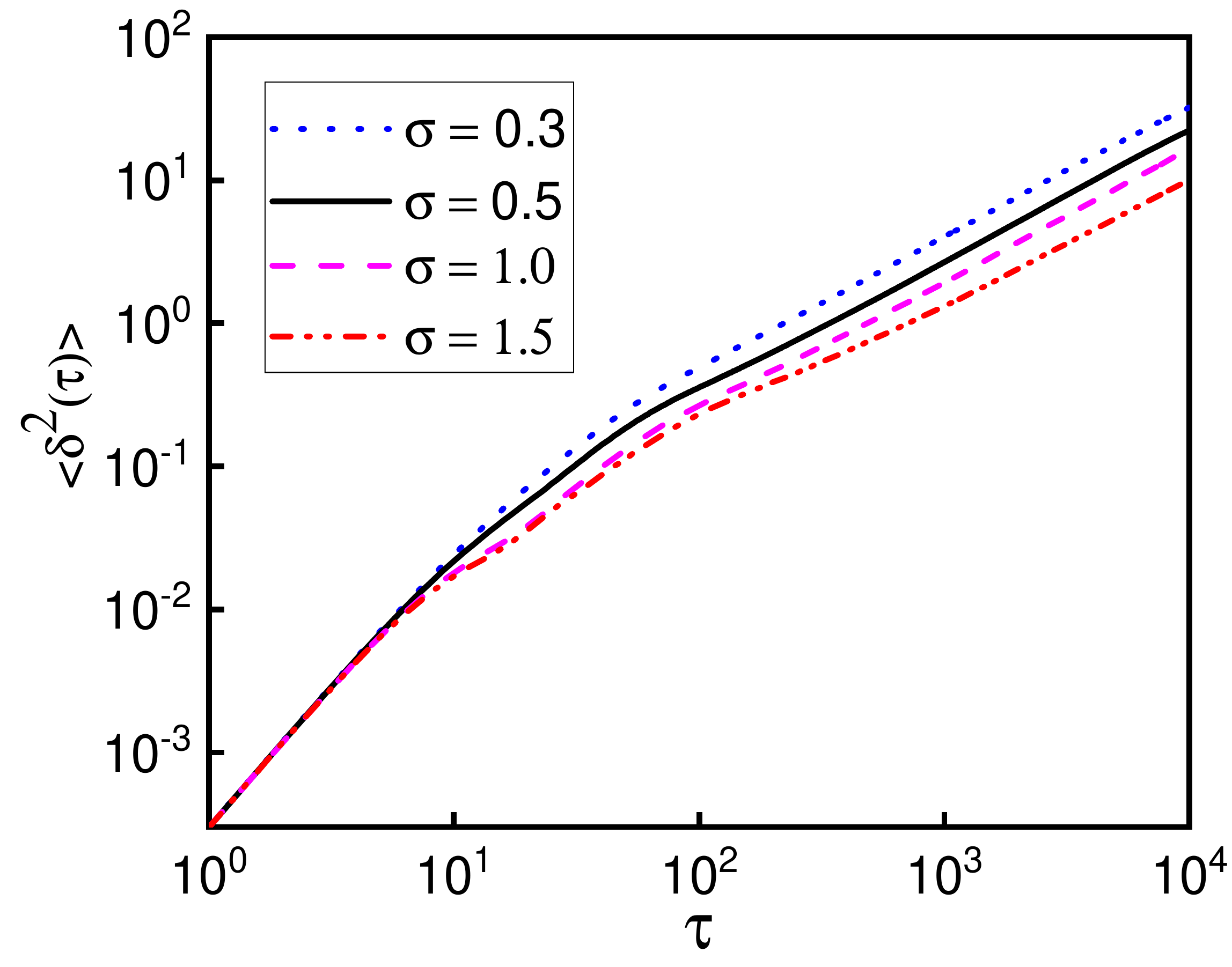} & 
		\includegraphics[width=0.33\textwidth]{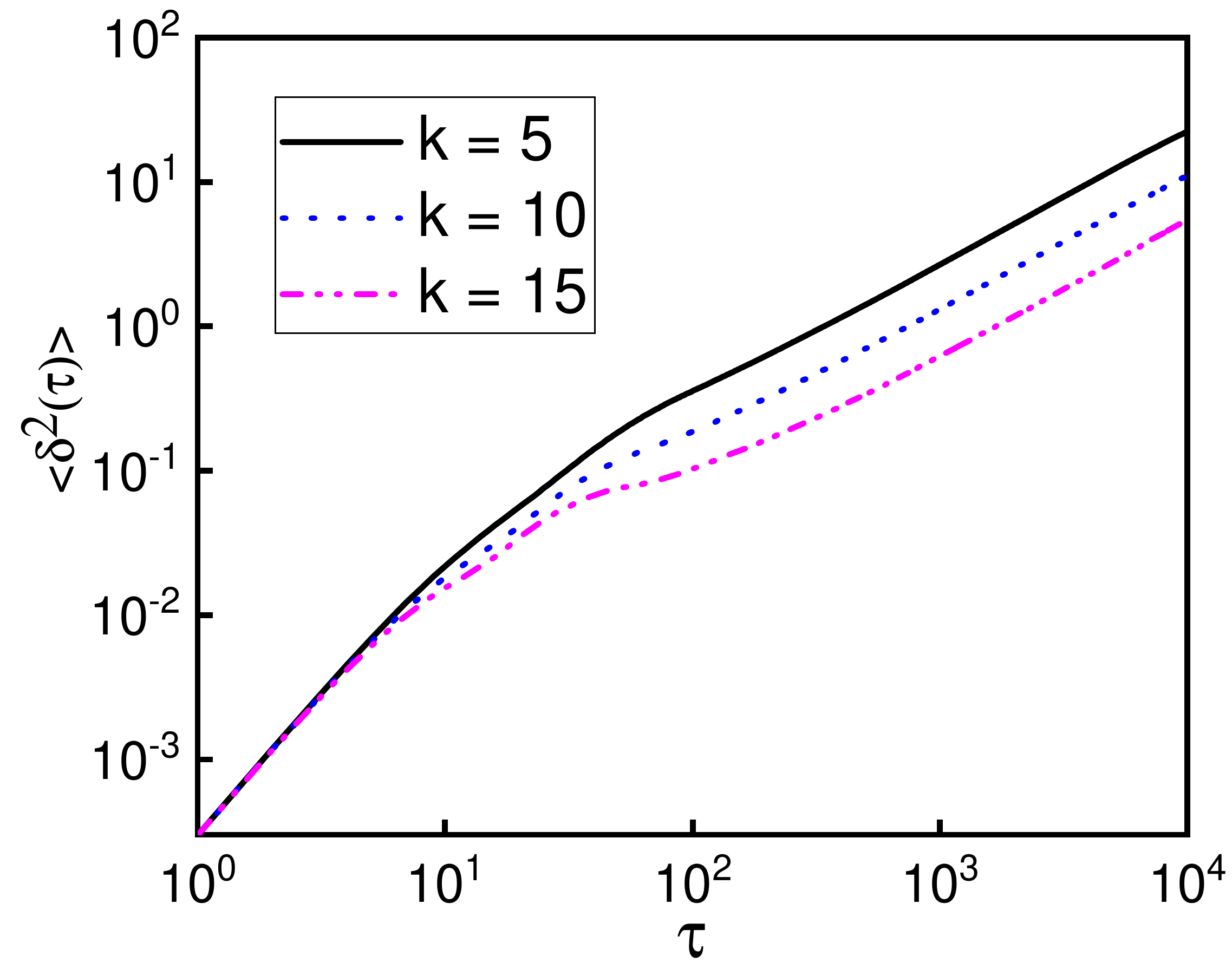}  \\

        (a) & (b) & (c) \\
		
		\includegraphics[width=0.33\textwidth]{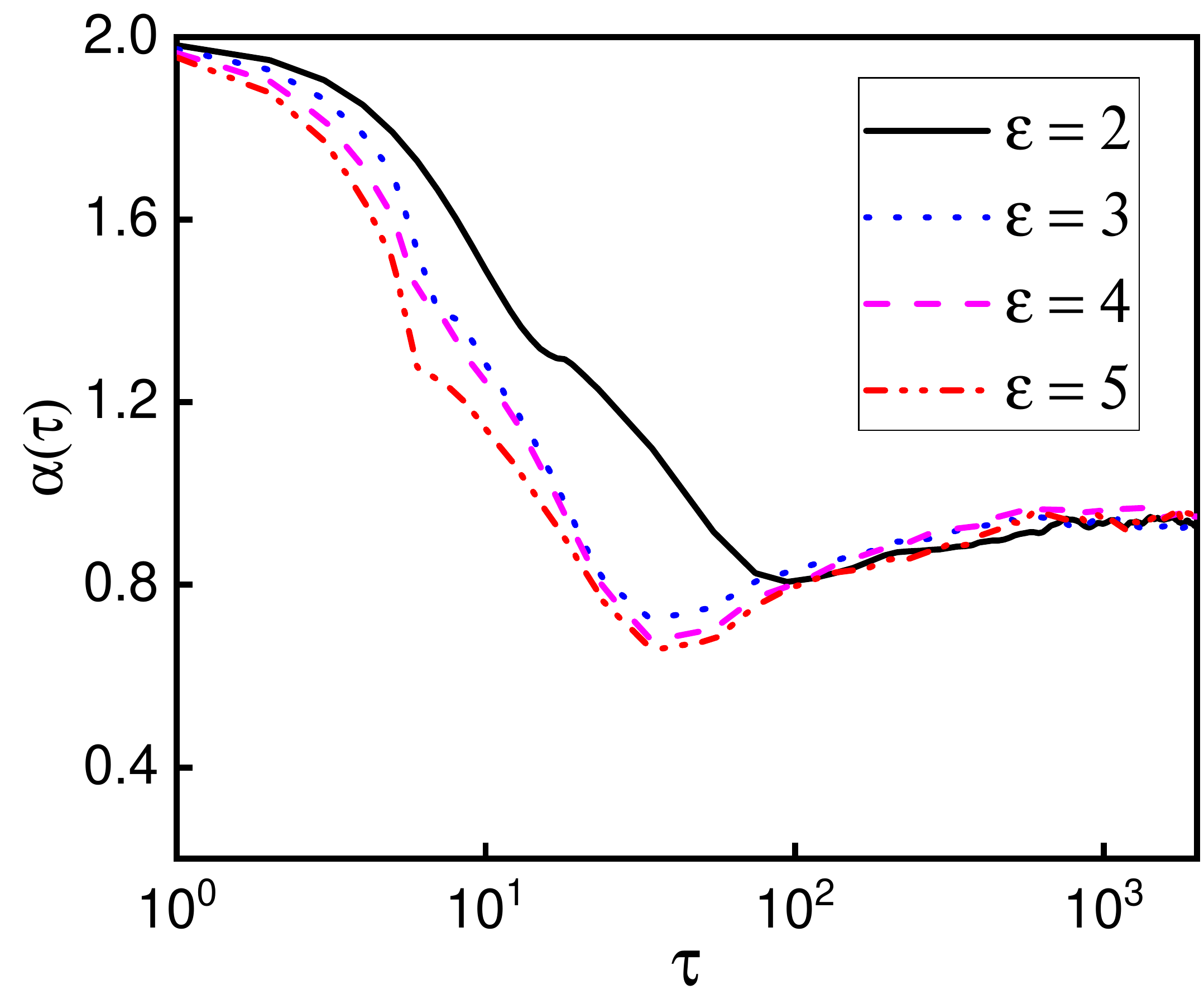} &
		\includegraphics[width=0.33\textwidth]{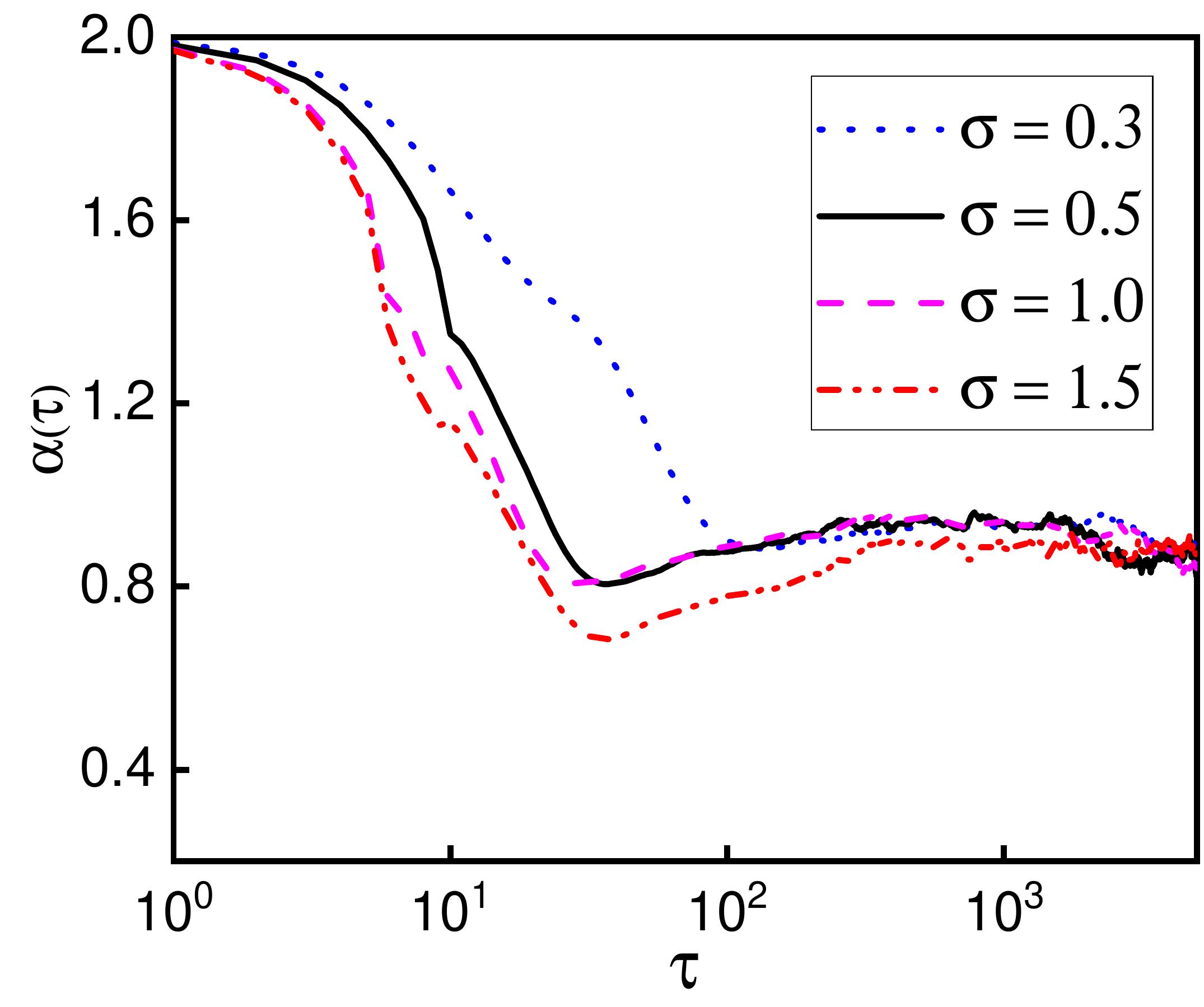} & 
		\includegraphics[width=0.33\textwidth]{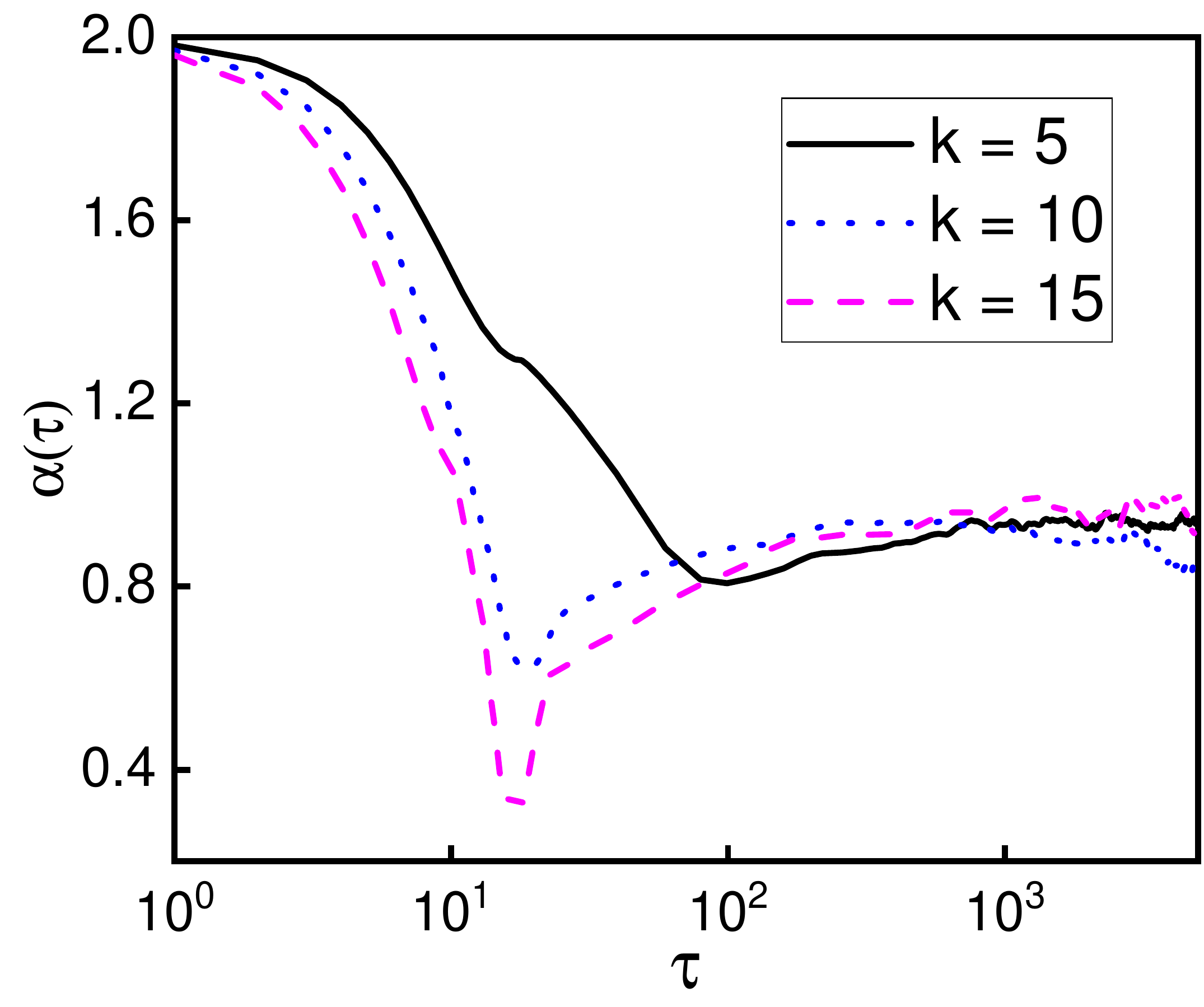} \\
        
        (d) & (e) & (f) \\	
		
		\includegraphics[width=0.33\textwidth]{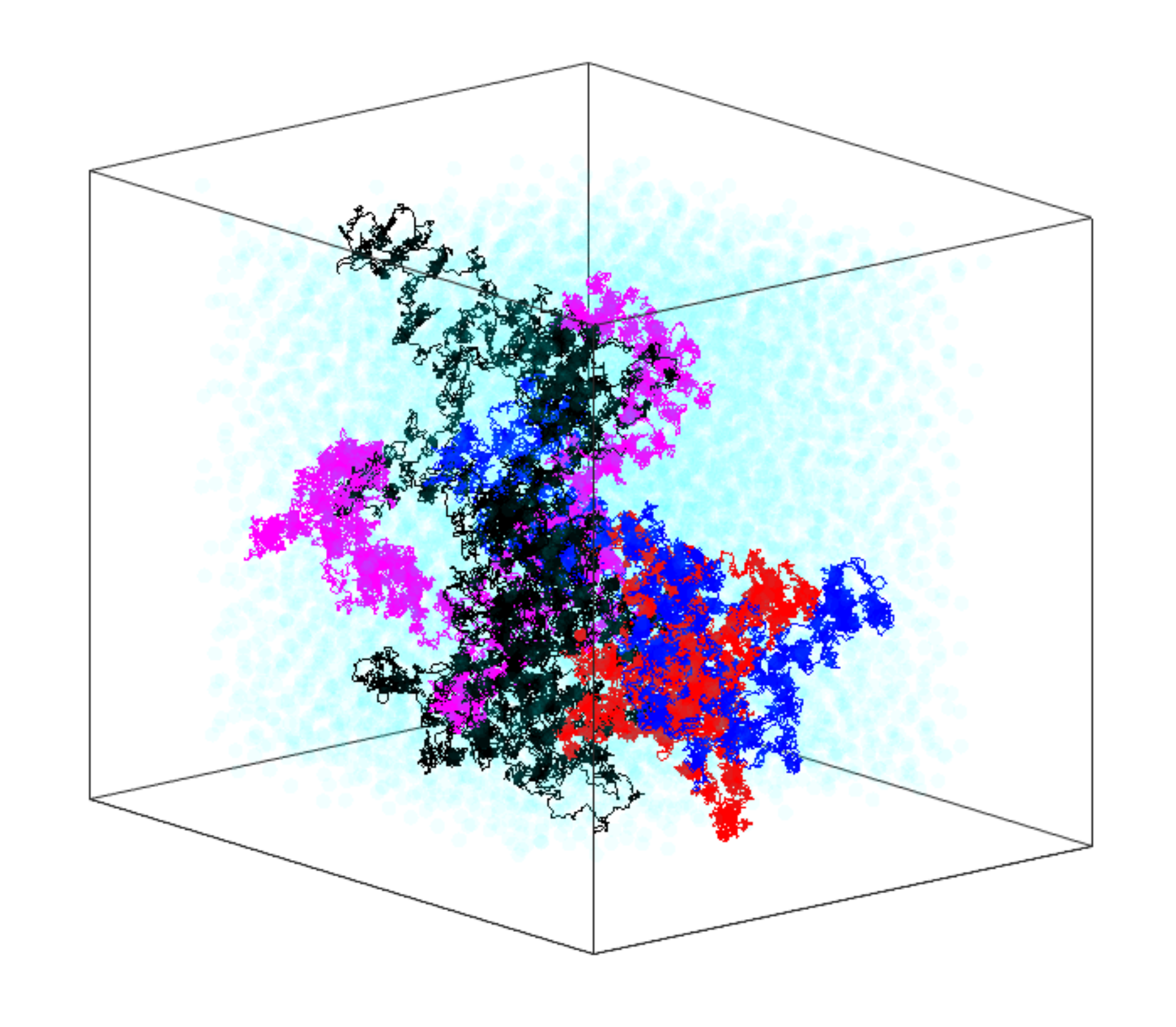} &
		\includegraphics[width=0.33\textwidth]{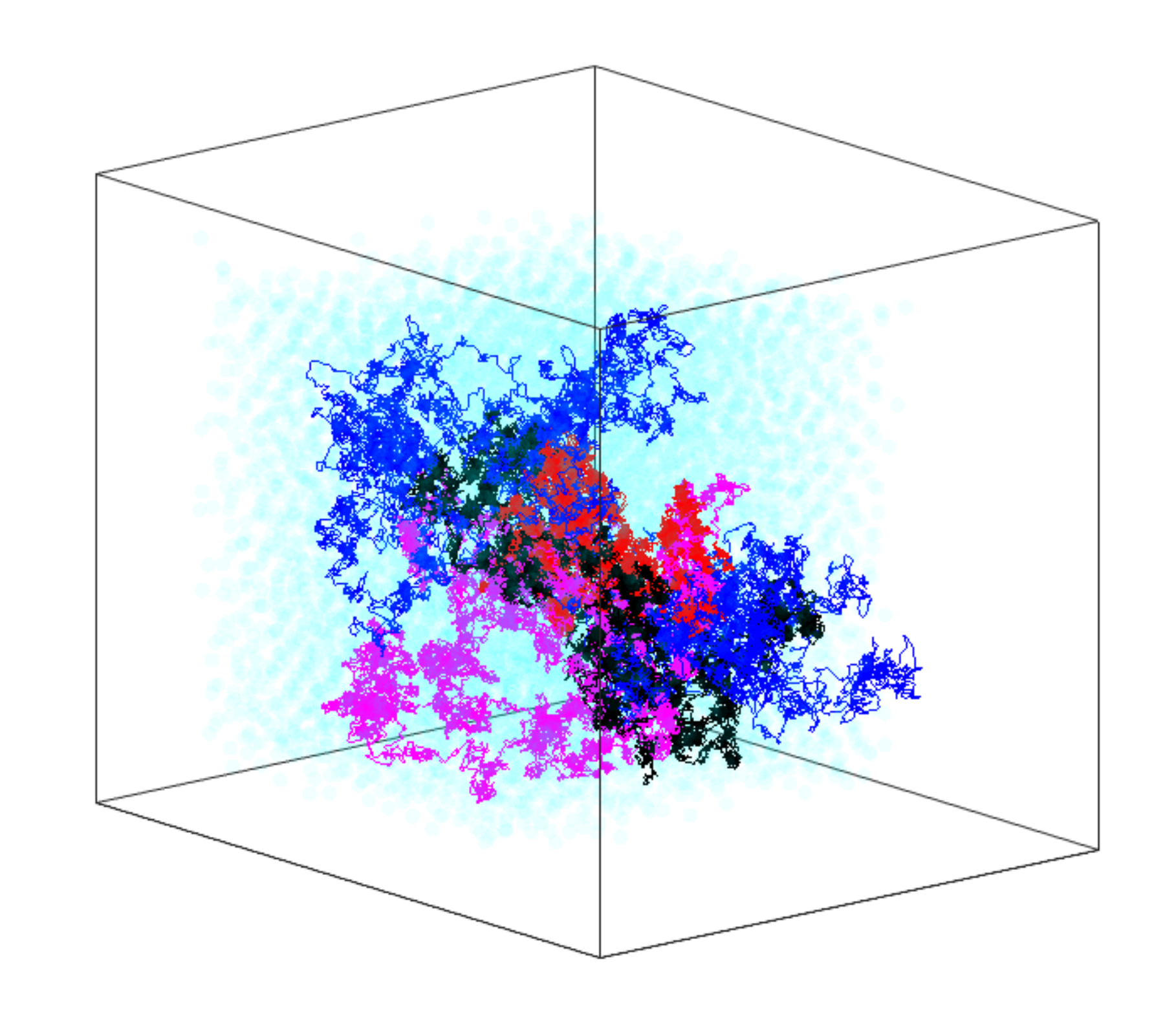} & 
		\includegraphics[width=0.33\textwidth]{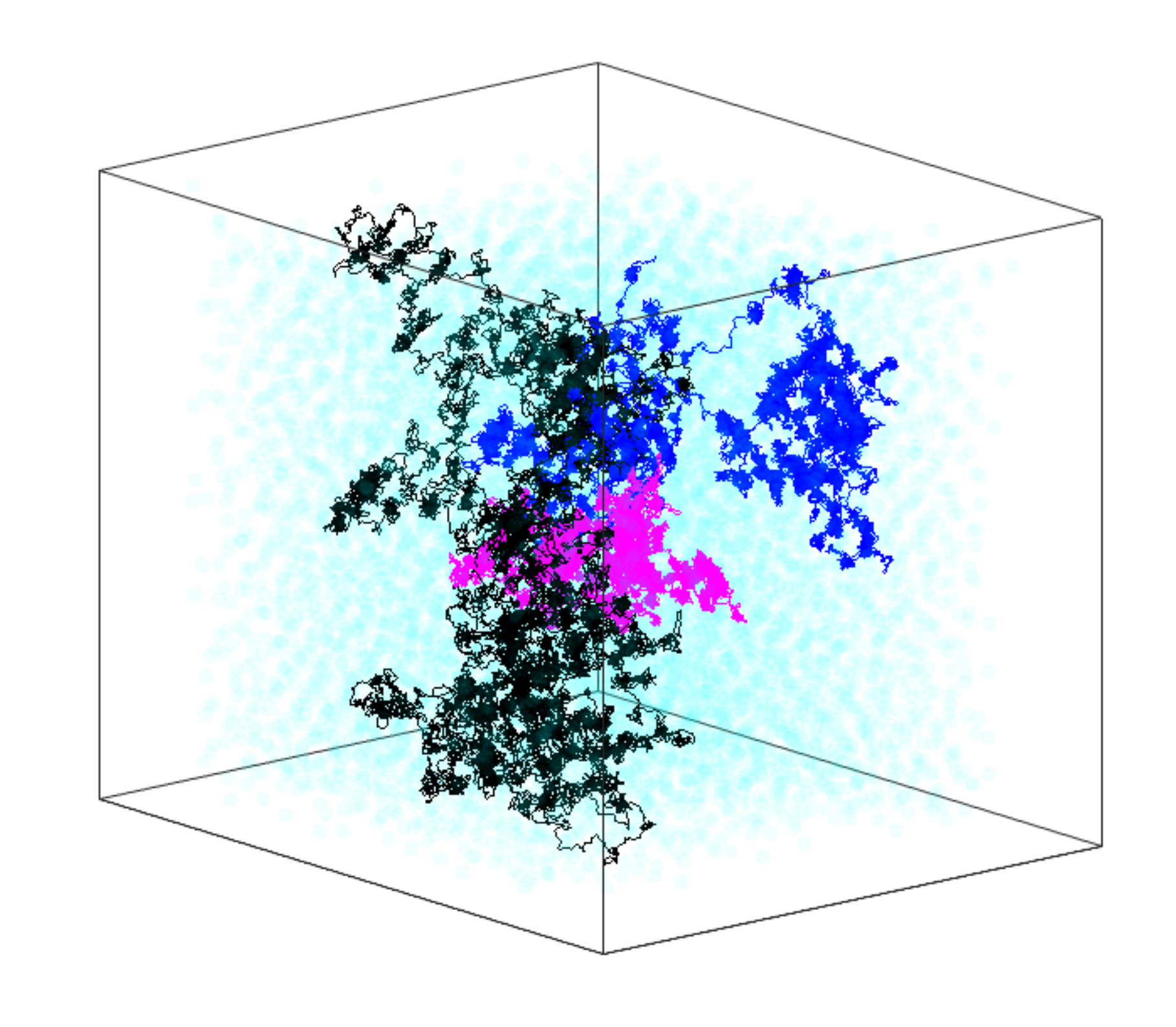} \\
		 (g) & (h) & (i) 
	\end{tabular}
\caption{Plots of $\left<\overline{\delta^{2}(\tau)}\right>$ vs $\tau$ and the corresponding exponents ($\alpha$) for different tracer-monomer binding affinities $ (\epsilon)$ with $\sigma=0.5$, $k=5$ ((a), (d), (g)), different sizes of the tracer ($\sigma$) with $\epsilon=2$, $k=5$ ((b), (e), (h)) and the stiffness constants ($k$) with $\sigma=0.5$, $\epsilon=2$ ((c), (f), (i)). The single trajectory plots along with the network monomer positions (cyan)  at the beginning of the simulations are shown in (g) for different values of $\epsilon$, different values of $\sigma$ (h) and different values of $k$ (i). Color codes remain the same.}
\label{fig:msd}
\end{figure}

\noindent In order to study the influence of polymer network on the dynamics of nanoprobes, we consider the time-and-ensemble average of mean square displacement $\left(\left<\overline{\delta^{2}(\tau)}\right>\right)$ as a function of lag time $\tau$. First we compute the time-averaged MSD, $\overline{\delta^{2}(\tau)} = \frac{1}{T-\tau} \int_{0}^{T-\tau} {\left[ \textbf{r}(t+\tau) - \textbf{r}(t)\right]}^2  dt$, for all the initial time $t$ along the same trajectory. The ensemble average MSD $(\left\langle{\delta^{2}(\tau)}\right\rangle)$ is defined as the mean square displacement for each particle during time $\tau$ and then average over the entire ensemble (over independent trajectories) i.e $\left\langle{\delta^{2}(\tau)}\right\rangle  =  \frac{1}{N} \sum_{i=1}^{N}{\delta_{i}^{2}(\tau)}$. Finally, the time-and-ensemble-averaged MSD $\left( \left<\overline{\delta^{2}(\tau)}\right>\right) $ is obtained by performing double averaging, which means time averaging followed by ensemble averaging. For a given set of  parameters we generate $50$ trajectories of the tracer. Thus all our simulation results are averaged over $50$ trajectories, which means running $5$ independent simulations each with $10$ tracers.
\\
\\
Fig. \ref{fig:msd} (a), (b) and (c) depict the variation of MSD obtained by double averaging ($\left<\overline{\delta^{2}(\tau)}\right>$)  against the time difference $(\tau)$ for a range of binding affinities $(\epsilon)$, probe sizes $(\sigma)$ and chain stiffnesses $(k)$ in log-log scale. Fig. \ref{fig:msd} (d), (e) and (f) show variations for the corresponding time exponents $\alpha (\tau)$ defined as $\alpha (\tau)=\frac{d \left<\overline{\delta^{2}(\tau)}\right>}{d \log \tau}$ as a function of lag time $\tau$. For these parameters, both $\left<\overline{\delta^{2}(\tau)}\right>$ and $\alpha (\tau)$ clearly exhibit three distinct regimes $-$ short time ballistic regime $(\alpha \approx 2)$, intermediate subdiffusion $(\alpha < 1)$ and long time nearly free diffusion $(\alpha \approx 1)$. Because, at very short time, the motion of the tracer is not affected by the polymer network and it moves ballistically (since we simulate an underdamped Langevin equation Eq. (\ref{eq:langevineq})). As the time progresses, the tracer starts feeling the existence of the polymer network resulting an intermediate time slowing down of the motion of the probe particle. However, at long times, longer than the longest relaxation time of the network, the tracer performs nearly free diffusion.  \\
\\
With increasing $\sigma$,  $\left<\overline{\delta^{2}(\tau)}\right>$ of the particle grows slower reflecting the efficient caging and restriction of escaping of the bigger probe particle \cite{dell2014theory}. Particles with a smaller size can pass through the cages formed by the polymer network easily, while the particles with relatively higher $\sigma$ are transiently trapped in these cages (see Supplementary Movie 1) and show intermediate time strong subdiffusive behavior ($\alpha<<1$). At longer lag time $\tau$, the tracer escapes out from the cages and eventually, the dynamics become diffusive.  Similar trends in  $\left<\overline{\delta^{2}(\tau)}\right>$ are also observed at different $\epsilon$ values. Higher the $\epsilon$, stronger the subdiffusive behavior and smaller the exponent $\alpha$. This happens since with increasing $\epsilon$, the tracer tends to bind with the gel particles for a longer duration which leads to a slow down of  $\left<\overline{\delta^{2}(\tau)}\right>$ (see Supplementary Movie 2).   In the case of varying $k$, keeping $\epsilon$ and $\sigma$ constant, the polymer gel (network) particles become less mobile and form nearly static cages at higher $k$ which suppresses the motion of the tracer (see Supplementary Movie 3 and 4). Thus, $\left<\overline{\delta^{2}(\tau)}\right>$ and $\alpha (\tau)$ exhibit strongly subdiffusive behavior with increasing $k$ as shown in Fig. (\ref{fig:msd}). In single particle tracking experiment with small organic probe molecules as the probes in polymer thin films, drying of the polymer films also have resulted similar stronger subdiffusion of the probe \cite{bhattacharya2013plasticization}. However, From Fig. (\ref{fig:msd}) we see that the effect of $k$ on $\left<\overline{\delta^{2}(\tau)}\right>$ and $\alpha (\tau)$  is more profound in comparison to $\epsilon$ and $\sigma$. We plot the trajectories of the tracer particle obtained from the simulation in Fig. (\ref{fig:msd}) and those are consistent with both $\left<\overline{\delta^{2}(\tau)}\right>$ and $\alpha (\tau)$. On increasing the chain stiffness ($k$), stickiness ($\epsilon$) or the probe size ($\sigma$) the trajectories become more localized confirming more restricted motion. These trajectories are like conformations of polymers and as $\epsilon$, $k$ or $\sigma$ increases localization of the trajectories can be viewed as conformations of polymers in poorer solvents \cite{flier2011heterogeneous}.
\\
\\
The effect of binding affinity and network stiffness on the dynamics of the probe is further characterized by calculating their long-time diffusivity $D^\alpha$. At longer time differences, we have reproduced the diffusivity ratio $\left(D^\alpha=\frac{\left<\overline{\delta^{2}(\tau)}\right>}{6\tau^{\alpha}}\right)$ for a range of $\alpha (\tau) (0.92-0.98)$ and the long-time diffusivity $D^\alpha$ will be the average over all these diffusivities. We tabulate $\frac{D^\alpha}{D_0}$ at different $\epsilon$ and $k$ for probe size $\sigma=0.5$ in Table (\ref{tab:table1}) where $D_0$ is the diffusivity of the free particle of same size obtained from independent simulations of the probe in absence of the network (simulation data not shown). A significant reduction of $D^\alpha$ has been observed in Table (\ref{tab:table1}) indicating the trapped motion of the probes in complex environment. Our reported diffusivities are in the same range as reported in the context of protein diffusing through model NPC central plug \cite{goodrich2018enhanced}.

\begin{table}
\begin{tabular}{llll}
\begin{tabular}{|c|c|c|c|c|}
\hline
$\epsilon$ & 2 & 3 & 4 & 5 \\
\hline
$\frac{D^{\alpha}}{D_{0}}$ & 0.0633 & 0.0421 & 0.0345 & 0.0272 \\
\hline
\end{tabular}
& \hspace{20mm}
\begin{tabular}{|c|c|c|c|}
\hline
$k$ & 5 & 10 & 15 \\
\hline
$\frac{D^{\alpha}}{D_{0}}$ & 0.0633 & 0.0303 & 0.0145 \\
\hline
\end{tabular}
\end{tabular}
\caption{Diffusivity ratio $\left( \frac{D^{\alpha}}{D_{0}}\right) $ for different $\epsilon$ with $\sigma$ = 0.5, $k$ = 5 and for different $k$ with $\epsilon = 2$, $\sigma= 0.5$ with $\alpha=0.92-0.98$.}
\label{tab:table1}
\end{table}

\subsection{Velocity autocorrelation $(C_v (\tau))$}

\begin{figure}[h]
	\centering
	\begin{tabular}{ccc}
		\includegraphics[width=0.33\textwidth]{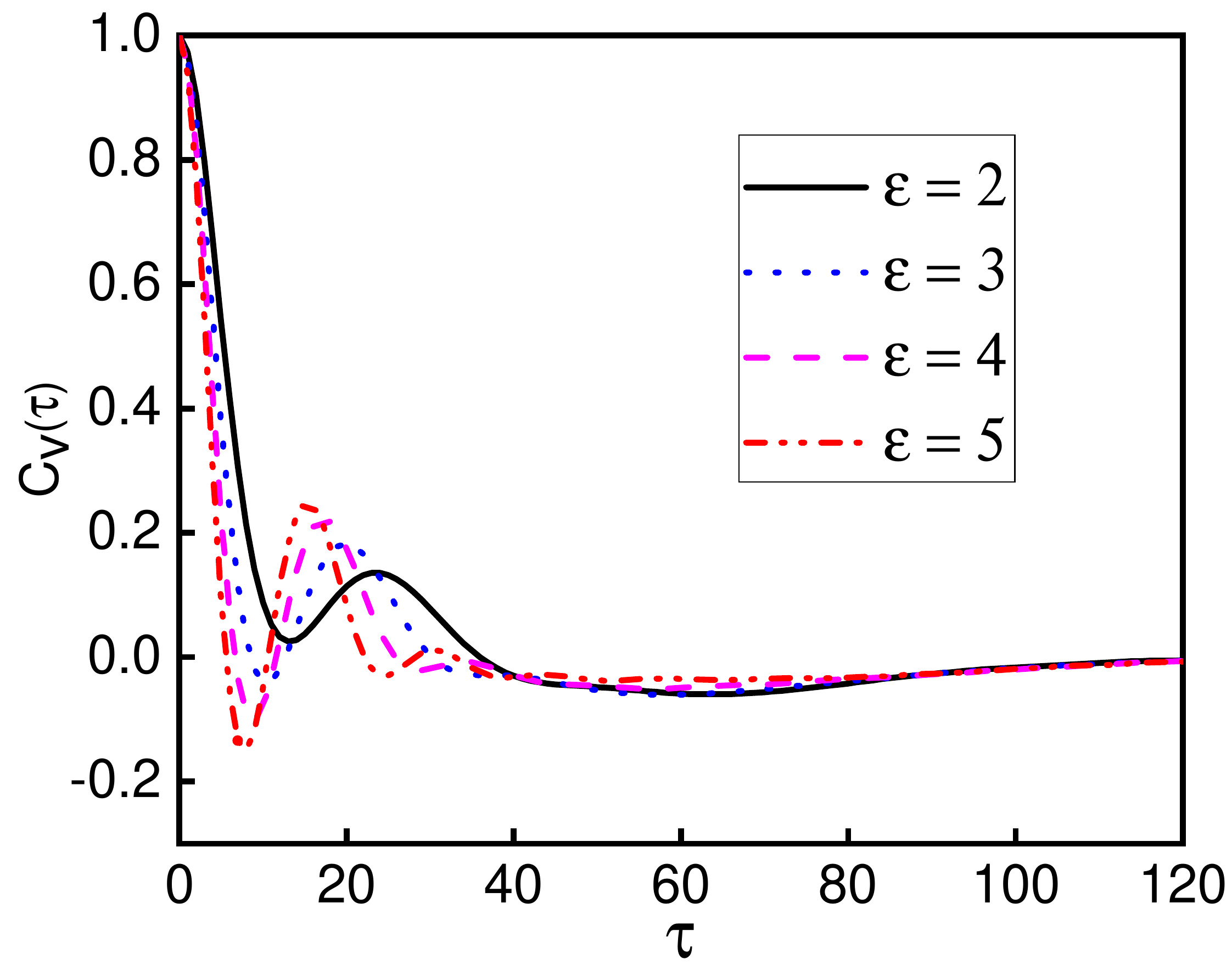} &
		\includegraphics[width=0.33\textwidth]{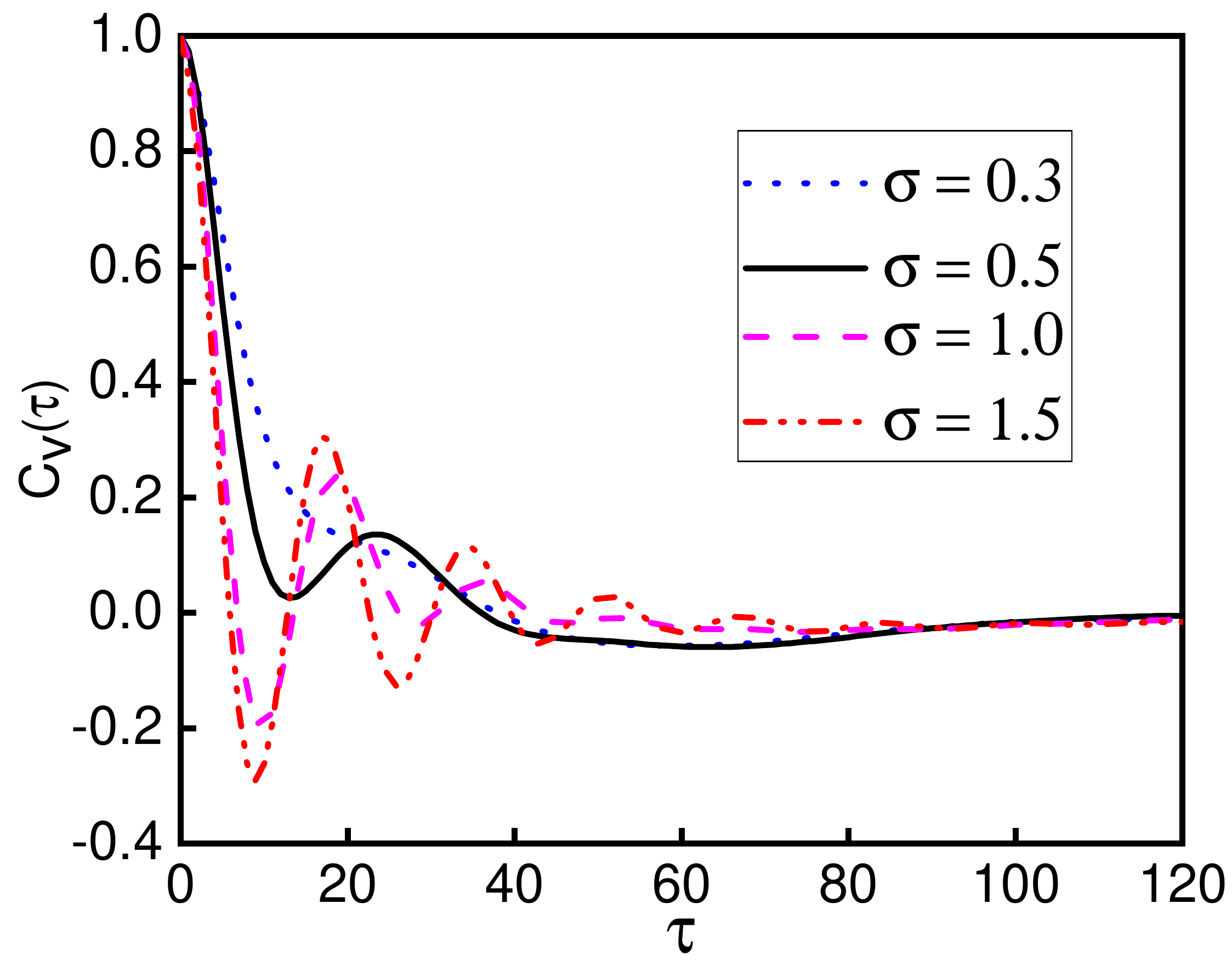} &
		\includegraphics[width=0.33\textwidth]{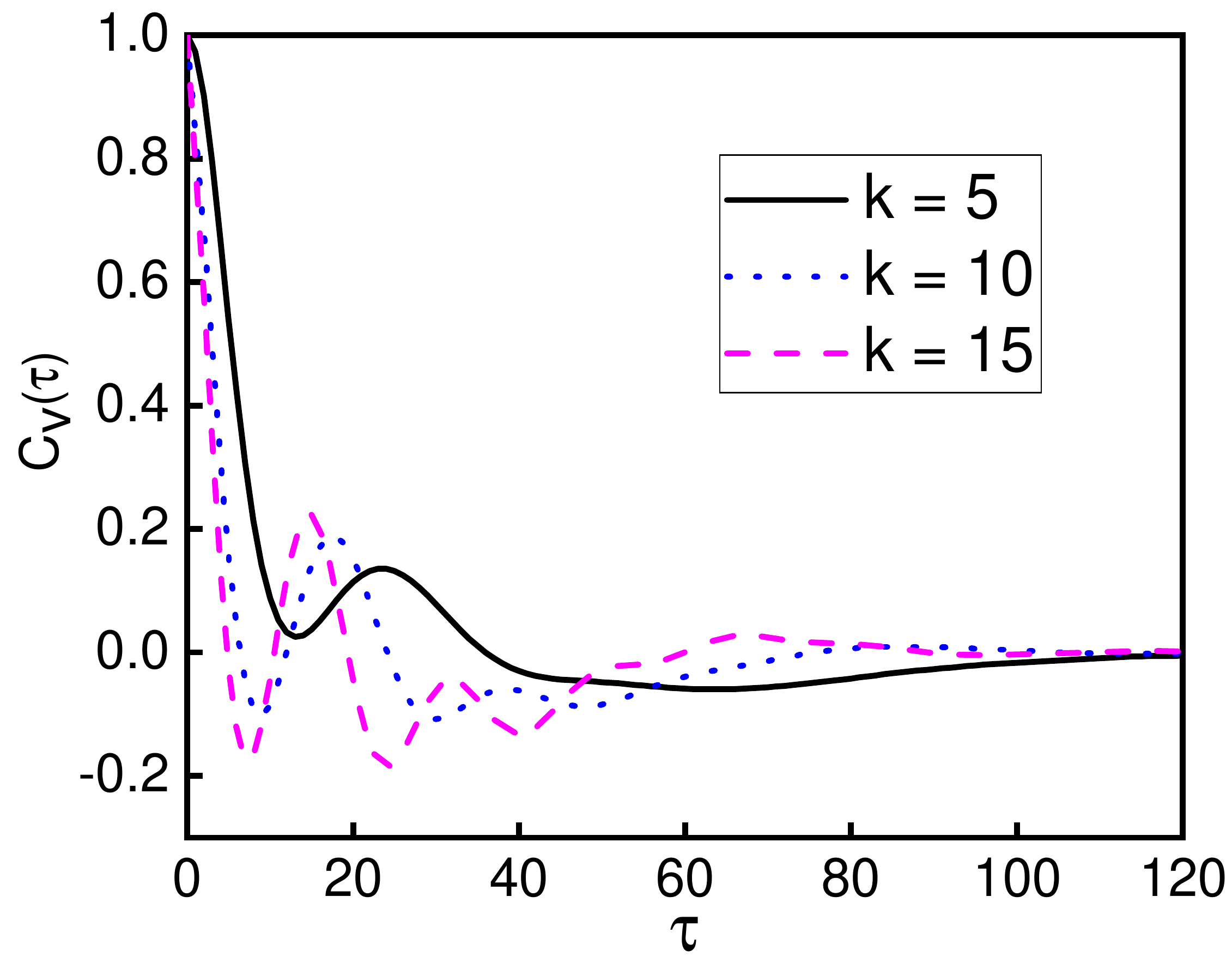} \\
		
		(a) & (b) & (c) \\
		
	\end{tabular}
	\caption{Plots of the VACF ($C_{v}(\tau)$) vs lag-time $\tau$ for different tracer-monomer binding affinities $ (\epsilon)$ with $\sigma=0.5$, $k=5$ ((a)), different sizes of the tracer ($\sigma$) with $\epsilon=2$, $k=5$ ((b)) and the stiffness constants ($k$) with $\sigma=0.5$, $\epsilon=2$((c)).}
	\label{fig:vacf}
\end{figure}

\noindent Other than MSD, in order to characterize the dynamics, especially what is happening at short time, we look at the velocity autocorrelation $(C_v (\tau))$, which is defined as $C_{v}(\tau) = \frac{\left\langle \overline{\textbf{v}(t+\tau).\textbf{v}(t)}\right\rangle}{\left\langle \overline{v^{2}(t)} \right\rangle}$. Plots of $C_v (\tau)$ \textit{vs} $\tau$ are shown in Fig.(\ref{fig:vacf}). In case of simple Brownian motion, $C_v (\tau)$ is always positive and decays exponentially to zero at longer lag time $\tau$, when the motions become completely uncorrelated. However, a pronounced feature of the $C_v (\tau)$ plots in Fig.(\ref{fig:vacf}) for higher values of $\sigma$, $\epsilon$, $k$ are dips into negative values at short time, an indication of negative correlation in the probe motion.  Such negative correlations can emerge primarily from two different mechanisms: the first is FBM \cite{chakrabarti2012dynamics} and the second is confined CTRW. The negative dips in $C_v (\tau)$ are more with higher $\epsilon$ and $\sigma$. This can clearly be seen by comparing the black solid curve and the dash-dot red curve in Fig.(\ref{fig:vacf}) (a) corresponding to  $(C_v (\tau))$ for $\epsilon=2$ and $\epsilon=5$. This is a manifestation of the viscoelastic response of the medium. For higher $\epsilon$ and $\sigma$, the probe attaches with the polymer network and moves back and forth following the motion of the polymer network. As a result, the motion of the probe in one direction is likely to be followed by the motion in opposite direction. Thus, FBM in viscoelastic medium emerges as the dominant mechanism for the motion of the probe. However, for higher $k$, the network will become rigid and form static cages. The probes are confined within these cages and doing jiggling motion. The probe collides with the polymer chains and scatters back within the cages. This leads to the confined CTRW type motion. Hence, we can get an intuitive picture about the dynamics of the probe in polymer gel. In general, CTRW and FBM both operate at the same time and account for the trapped motion that lead to negative velocity autocorrelation at short time \cite{samanta2016tracer} but as the polymers become more rigid the contributions from confined CTRW dominate.

\subsection{van-Hove function}

\begin{figure}[h]
	\centering
	\begin{tabular}{ccc}
		\includegraphics[width=0.33\textwidth]{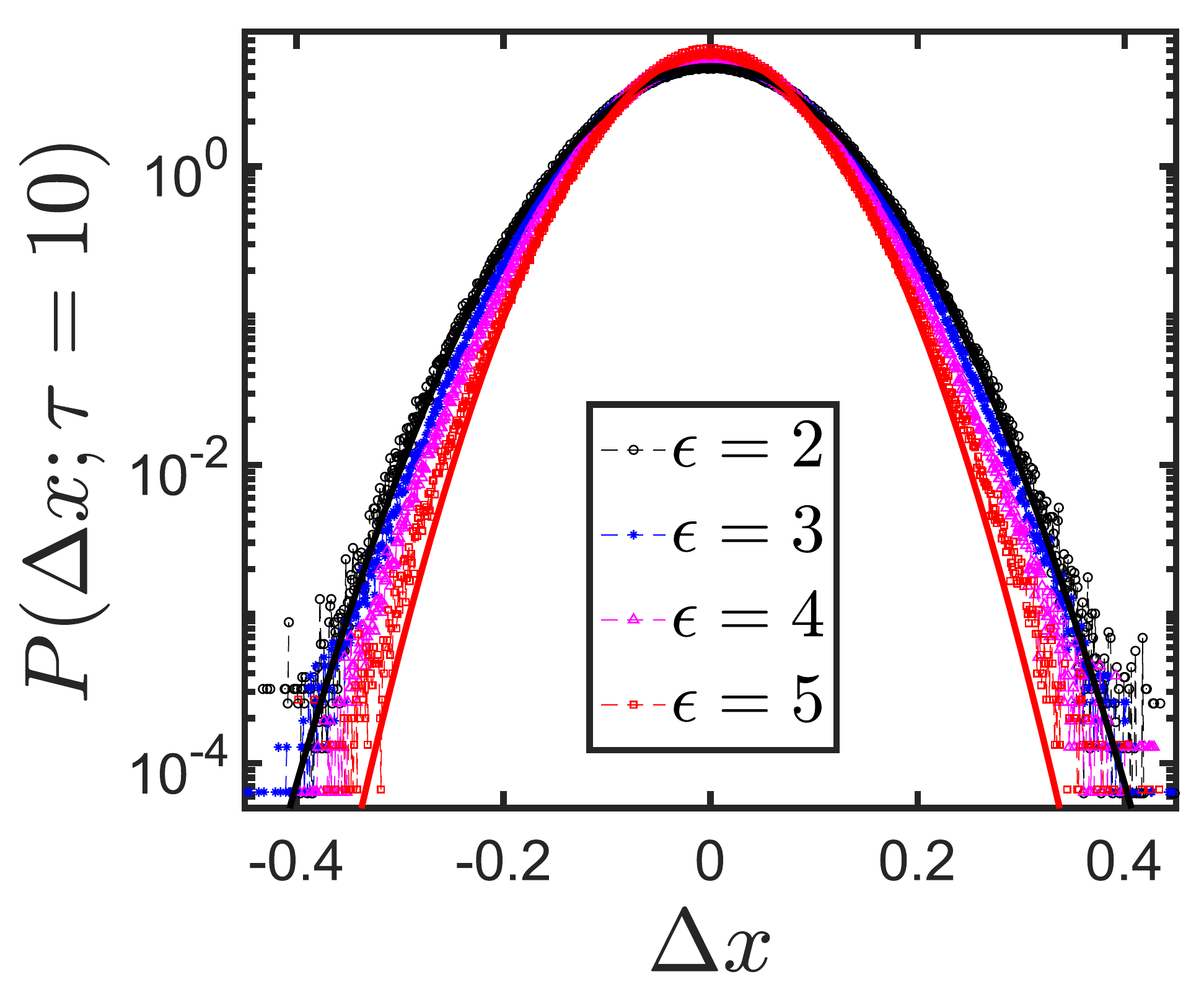}  &
		\includegraphics[width=0.33\textwidth]{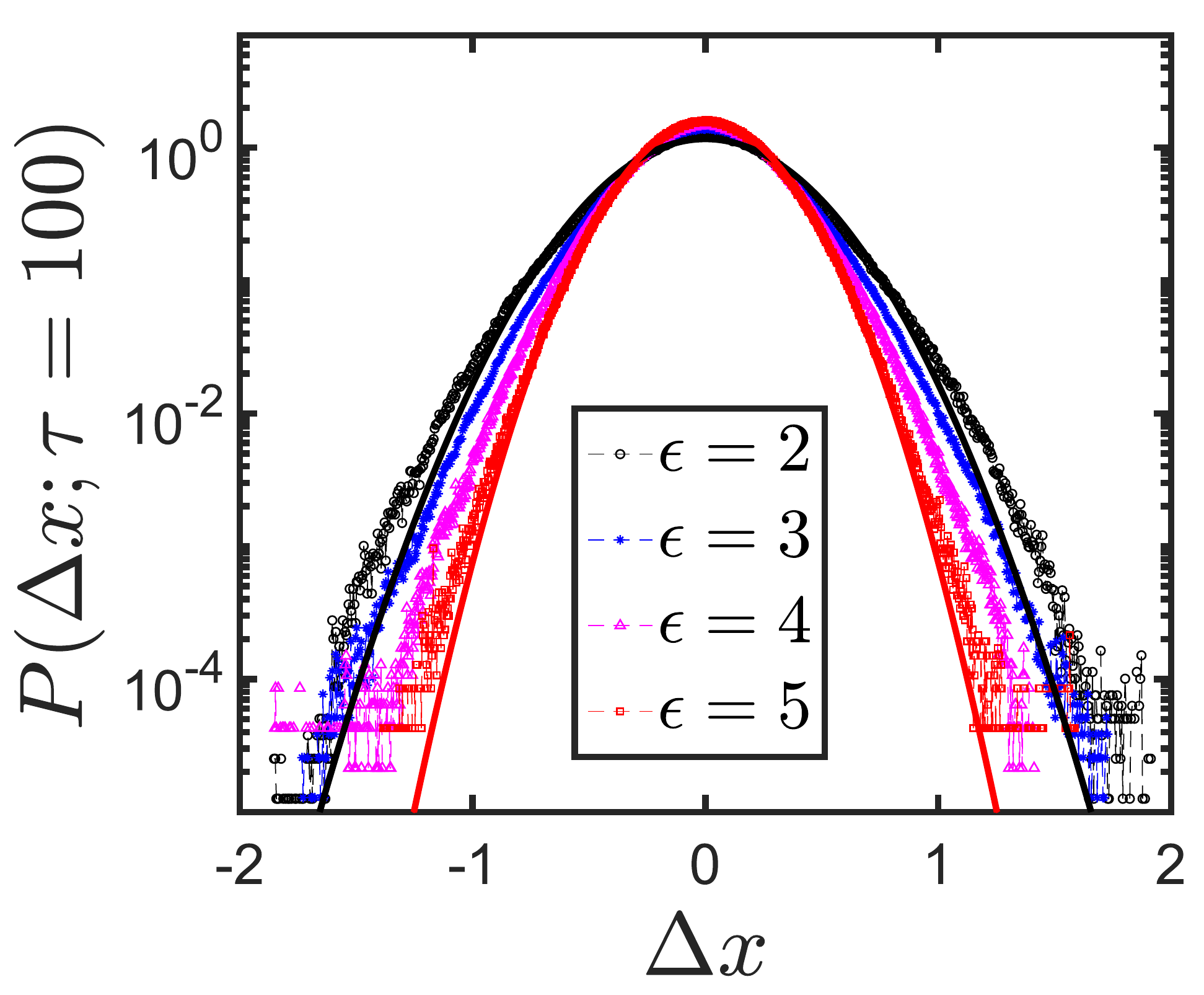} & 
		\includegraphics[width=0.33\textwidth]{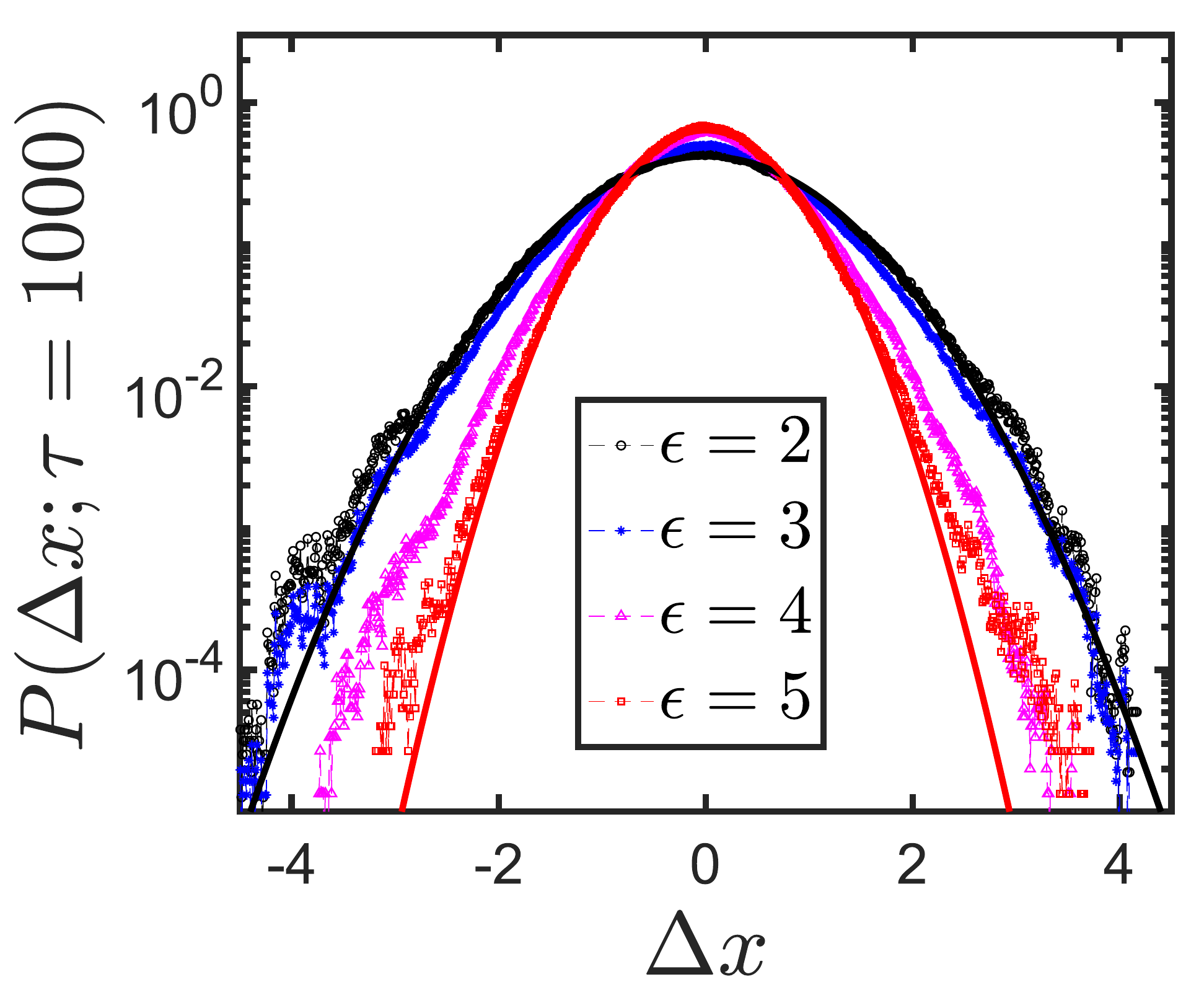}  \\
		
                      (a) & (b) & (c) \\		
		
		\includegraphics[width=0.33\textwidth]{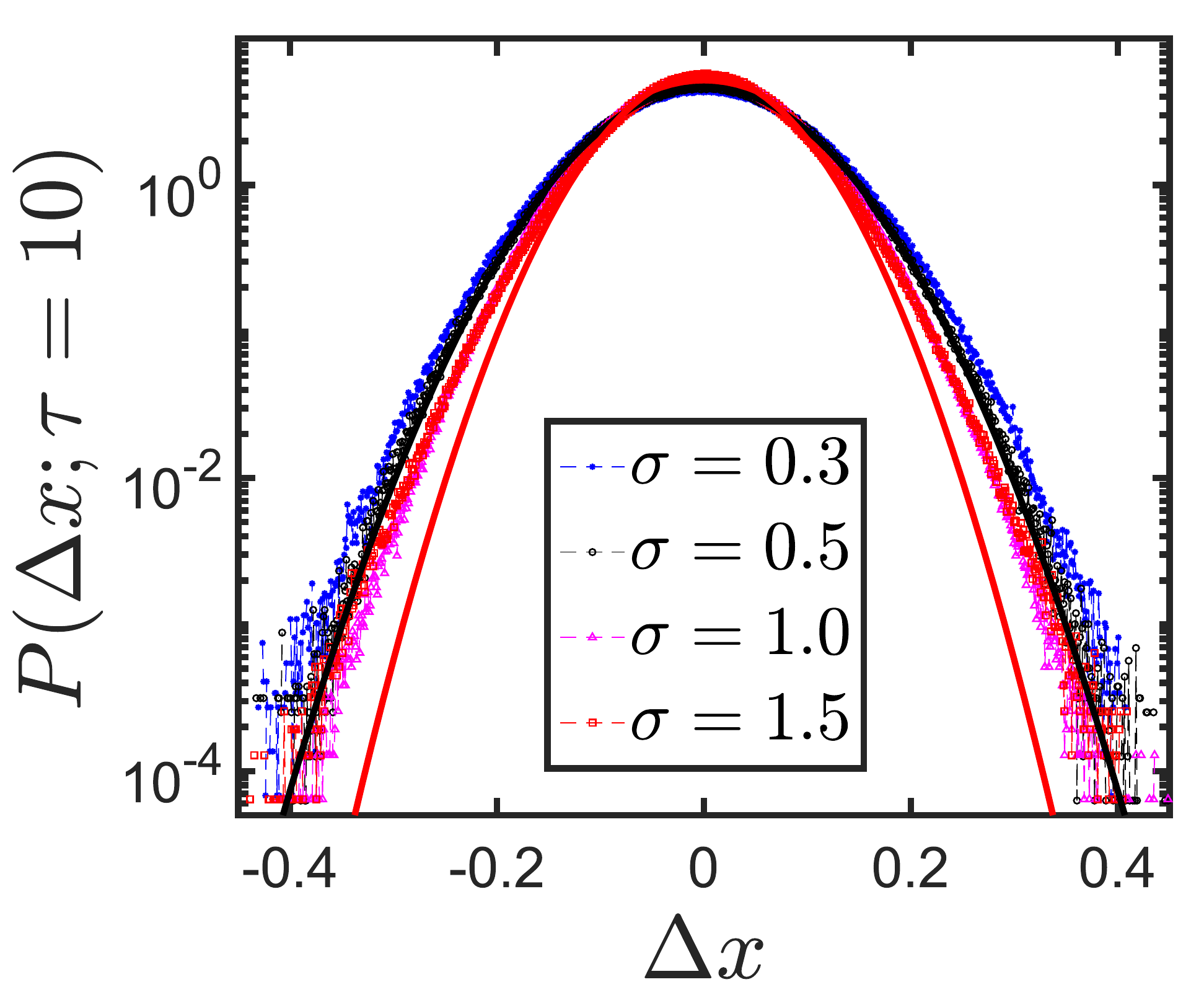} &
		\includegraphics[width=0.33\textwidth]{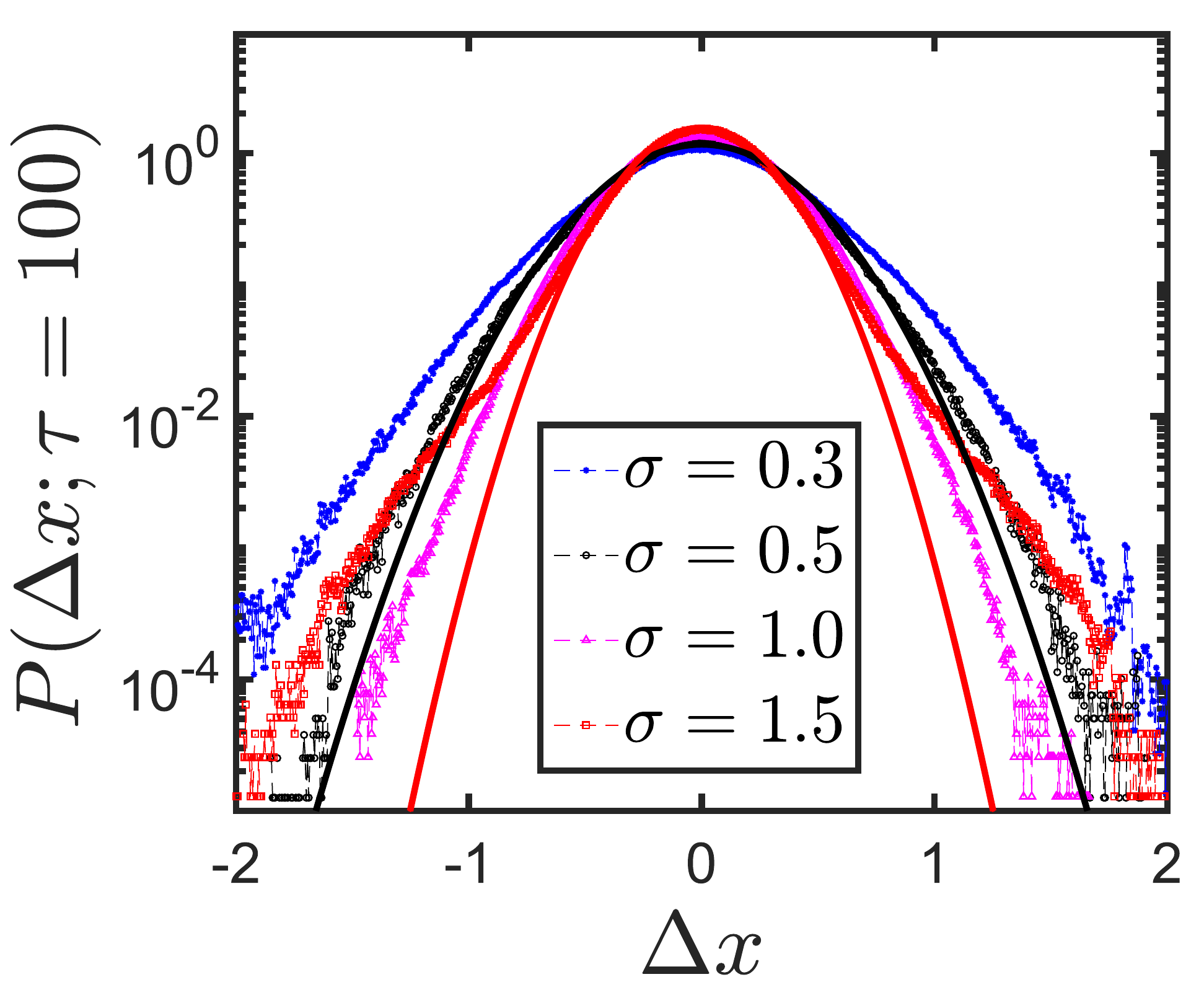} &  
		\includegraphics[width=0.33\textwidth]{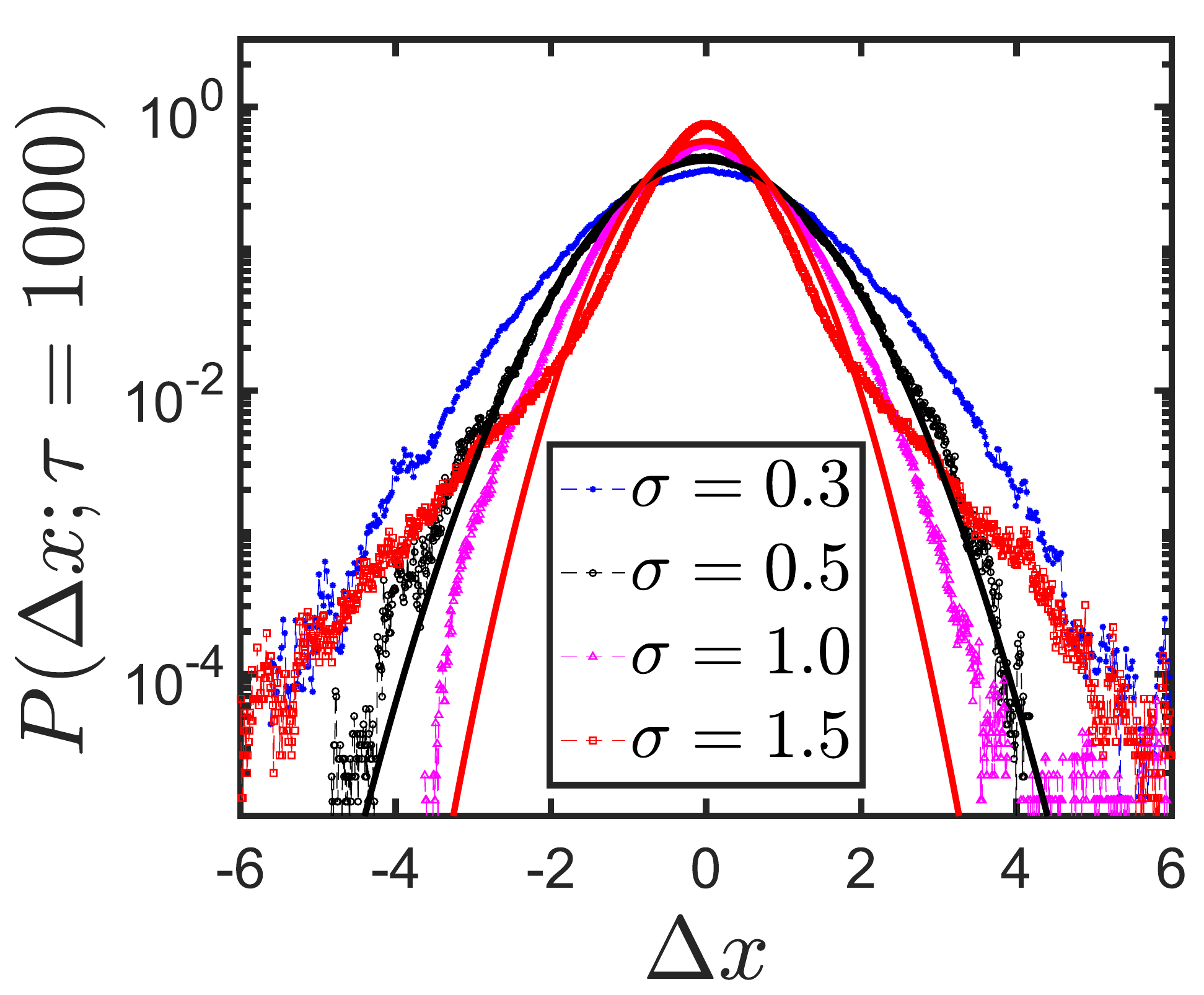} \\

                      (d) & (e) & (f) \\		
		
		\includegraphics[width=0.33\textwidth]{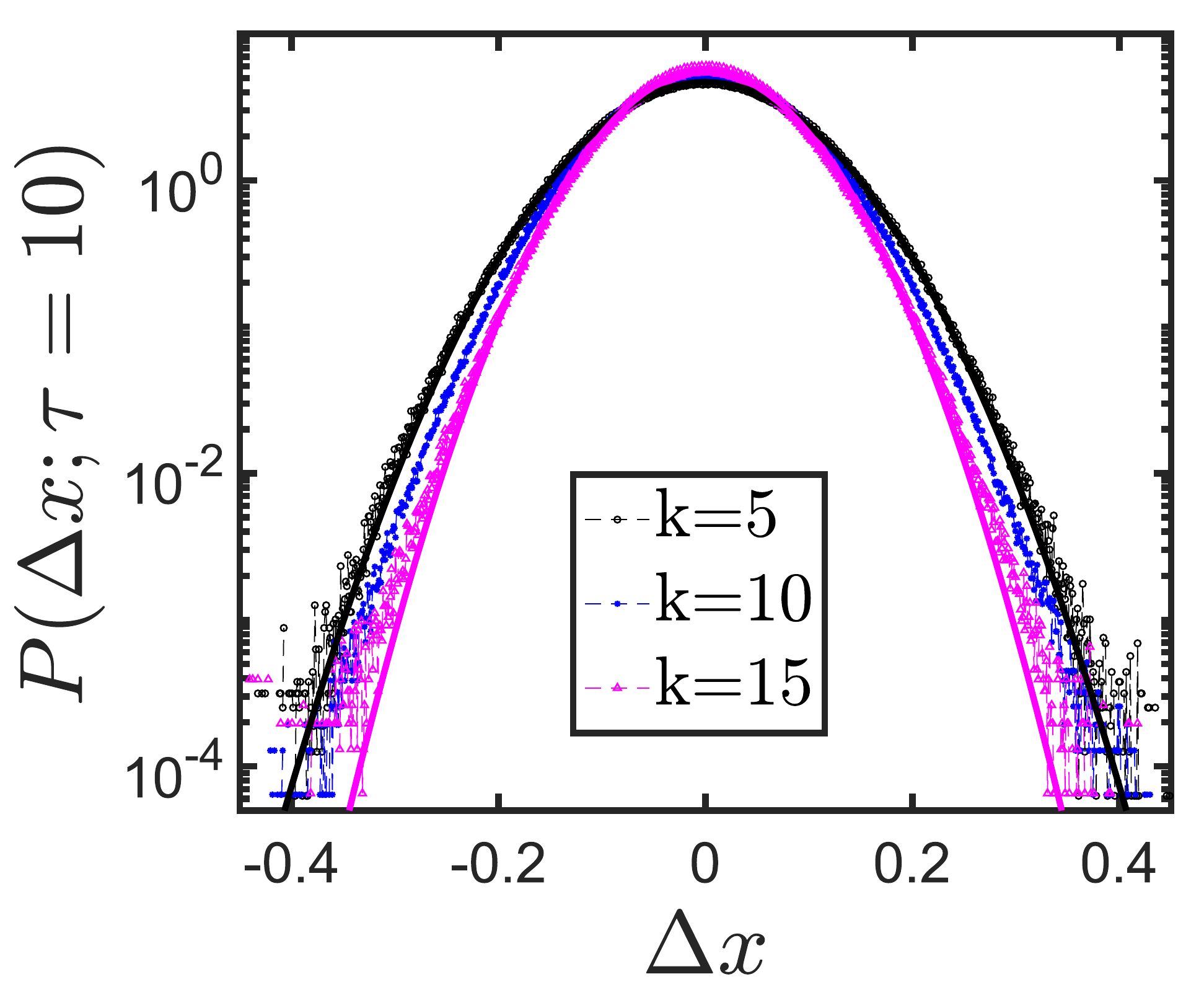} &
		\includegraphics[width=0.33\textwidth]{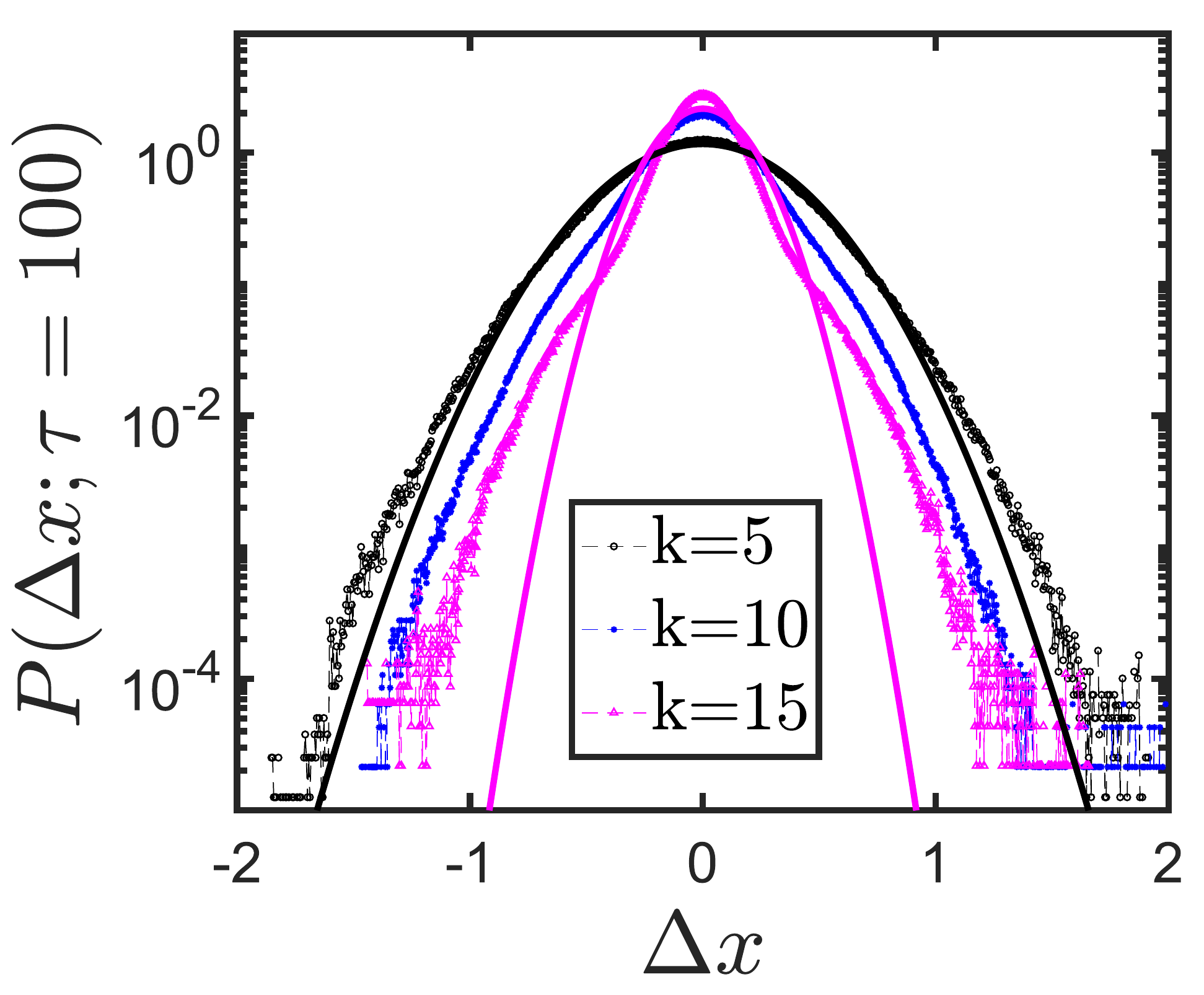} &  
		\includegraphics[width=0.33\textwidth]{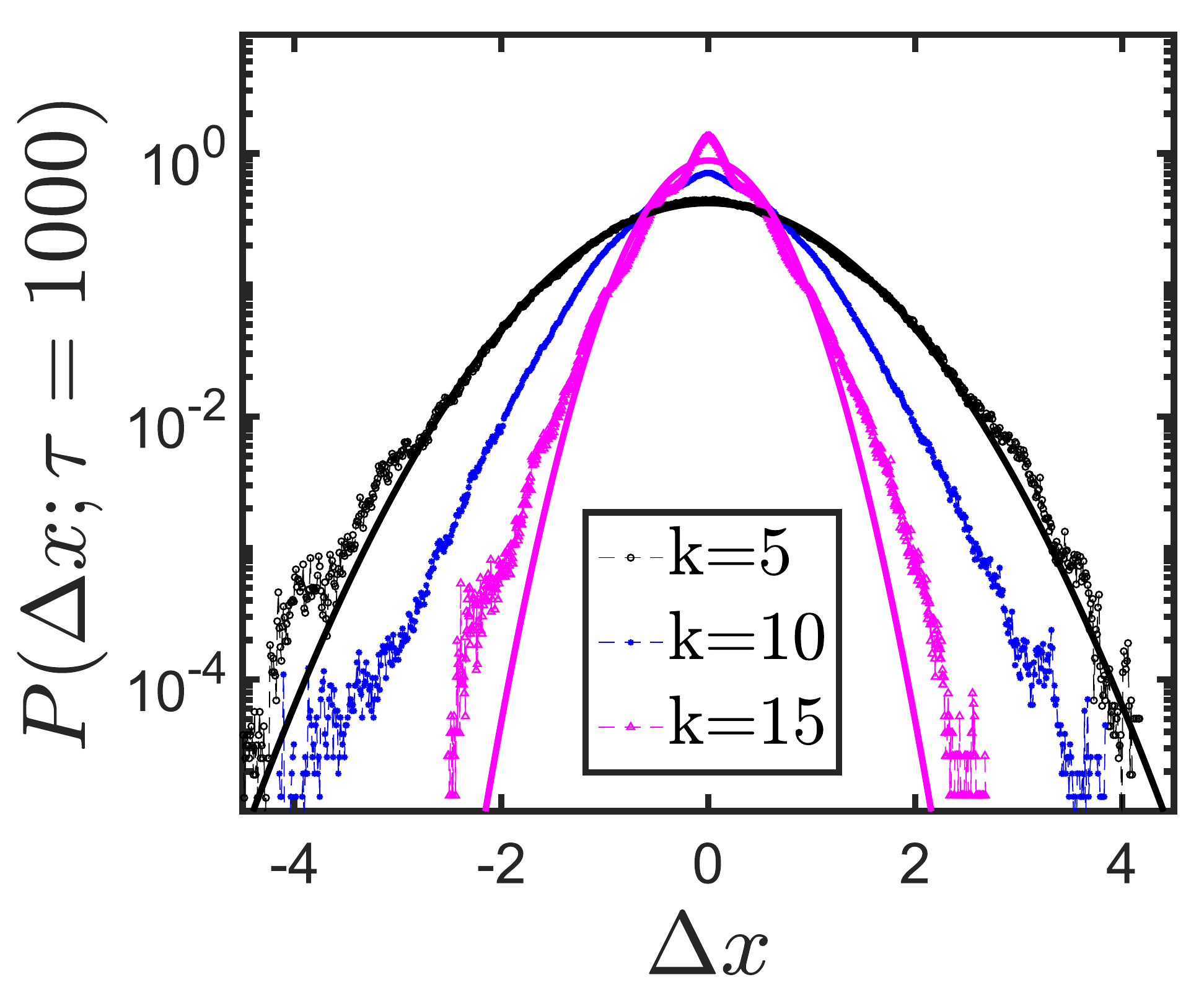} \\
		
                      (g) & (h) & (i) \\		
		
	\end{tabular}
\caption{Plots of the self part of the van-Hove functions in $x$, $P(\Delta x; \tau)$ at three different lag-times $\tau$ (10, 100, 1000)  for different tracer-monomer binding affinities $ (\epsilon)$ ((a), (b) and (c)) with $\sigma=0.5$, $k=5$, different sizes of the tracer ($\sigma$) ((d), (e) and (f)) with $\epsilon=2$, $k=5$ and the stiffness constants ($k$) ((g), (h) and (i)) with $\sigma=0.5$, $\epsilon=2$.}
\label{fig:vanhove}
\end{figure}

\begin{figure}[h]
	\centering
	\begin{tabular}{ccc}
		\includegraphics[width=0.33\textwidth]{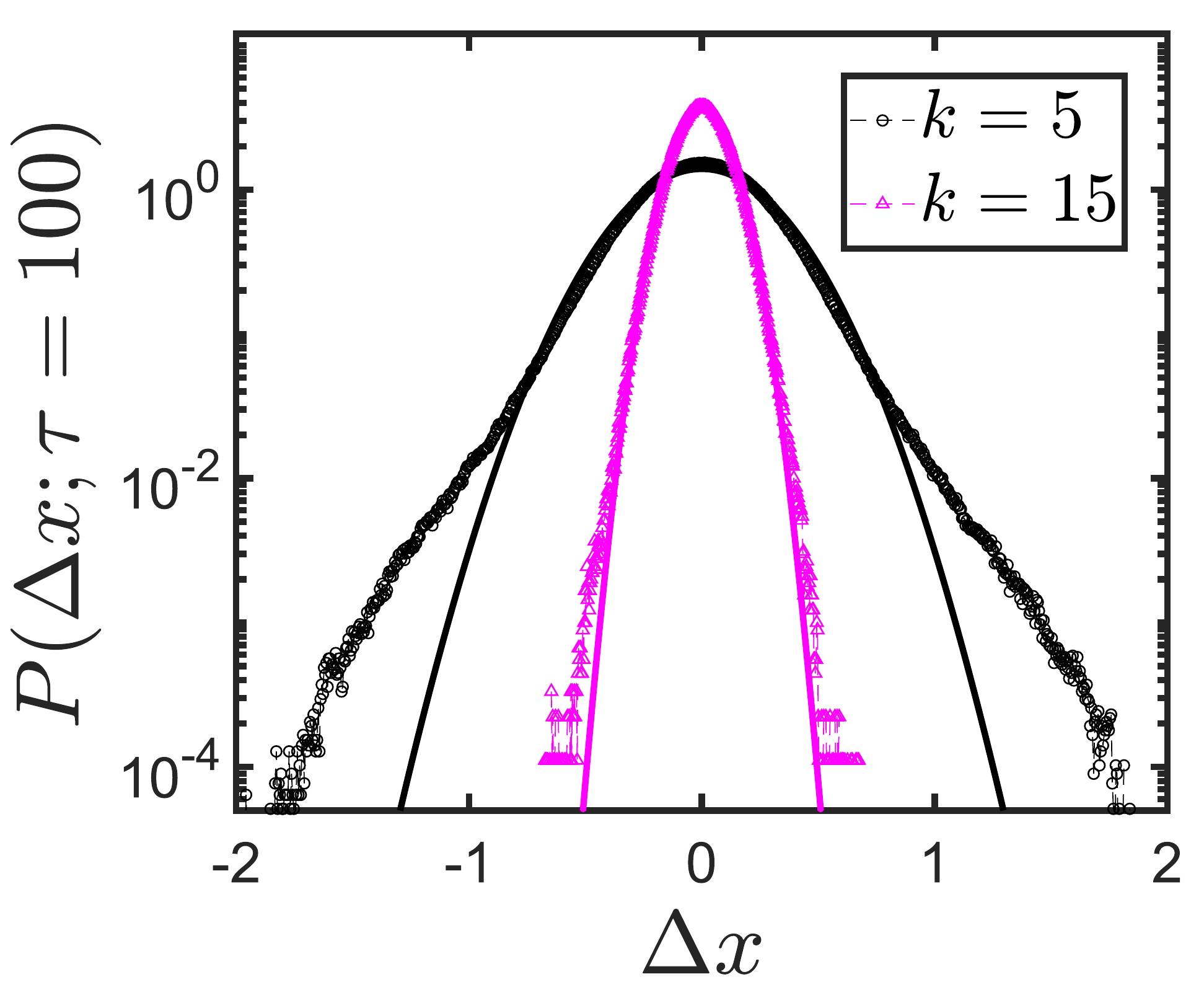} &
		\includegraphics[width=0.33\textwidth]{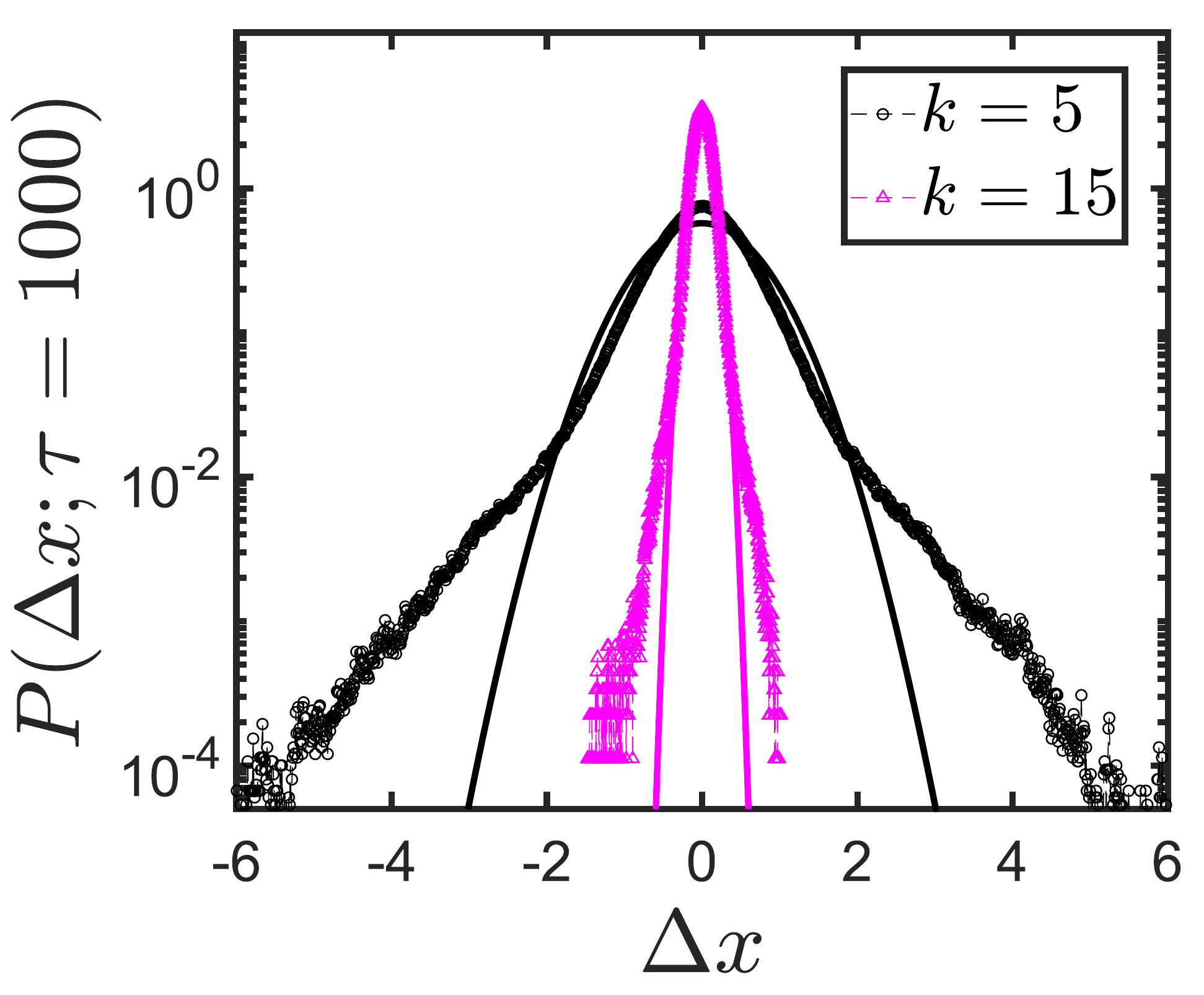} & 
		\includegraphics[width=0.33\textwidth]{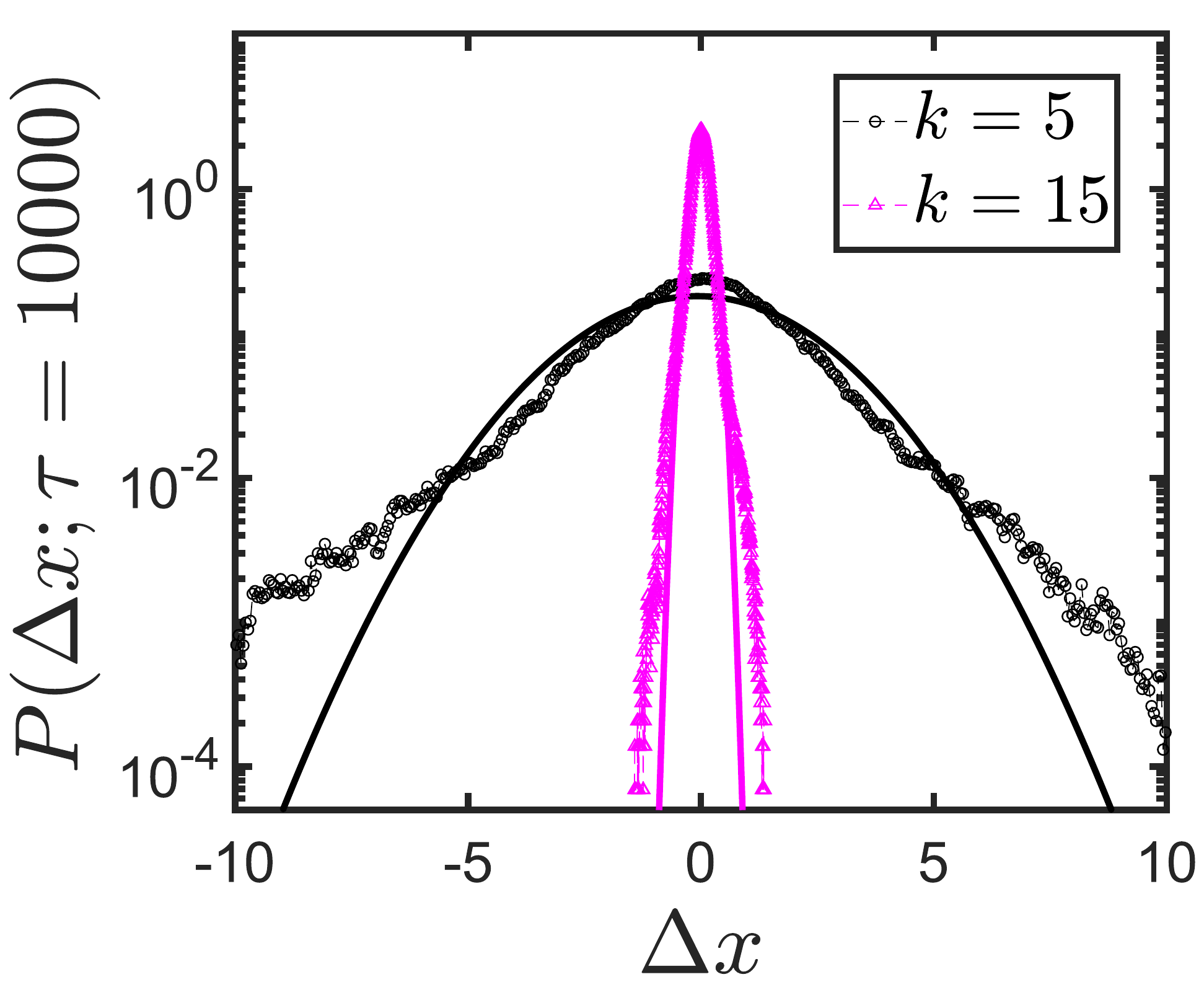} \\

                    (a) & (b) & (c) \\		
		
	\end{tabular}
\caption{Plots of the self part of the van-Hove functions in $x$, $P(\Delta x; \tau)$  for three chosen lag-times $\tau$ (100, 1000, 10000) with $\sigma=1.5$, $\epsilon=2$ and $k=5$ and $k=15$. Solid lines represent best Gaussian fits. }	
\label{fig:vanhovediffk}
\end{figure}

\begin{figure}[h]
	\centering
	\begin{tabular}{ccc}
		\includegraphics[width=0.33\textwidth]{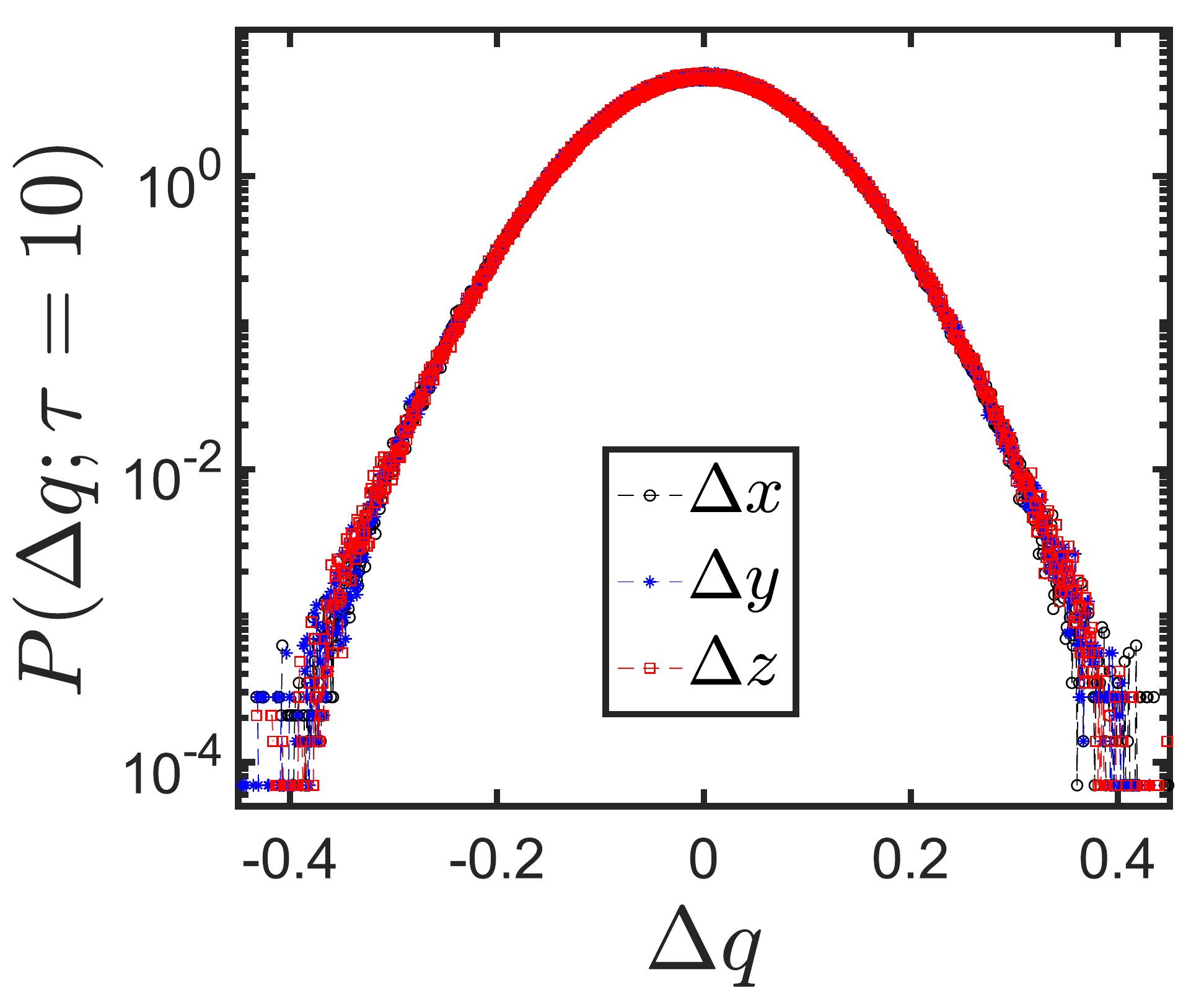} &
		\includegraphics[width=0.33\textwidth]{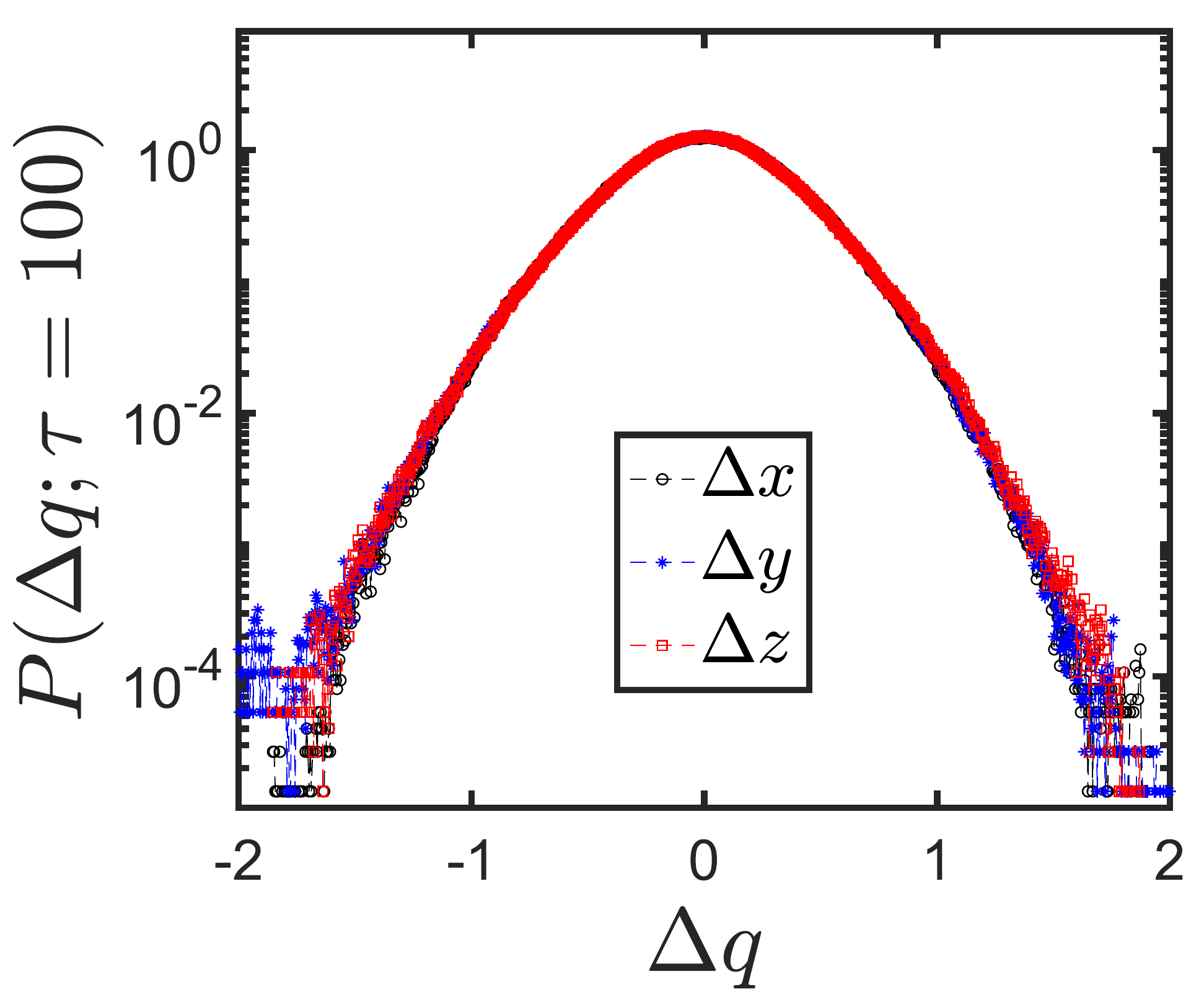} & 
		\includegraphics[width=0.33\textwidth]{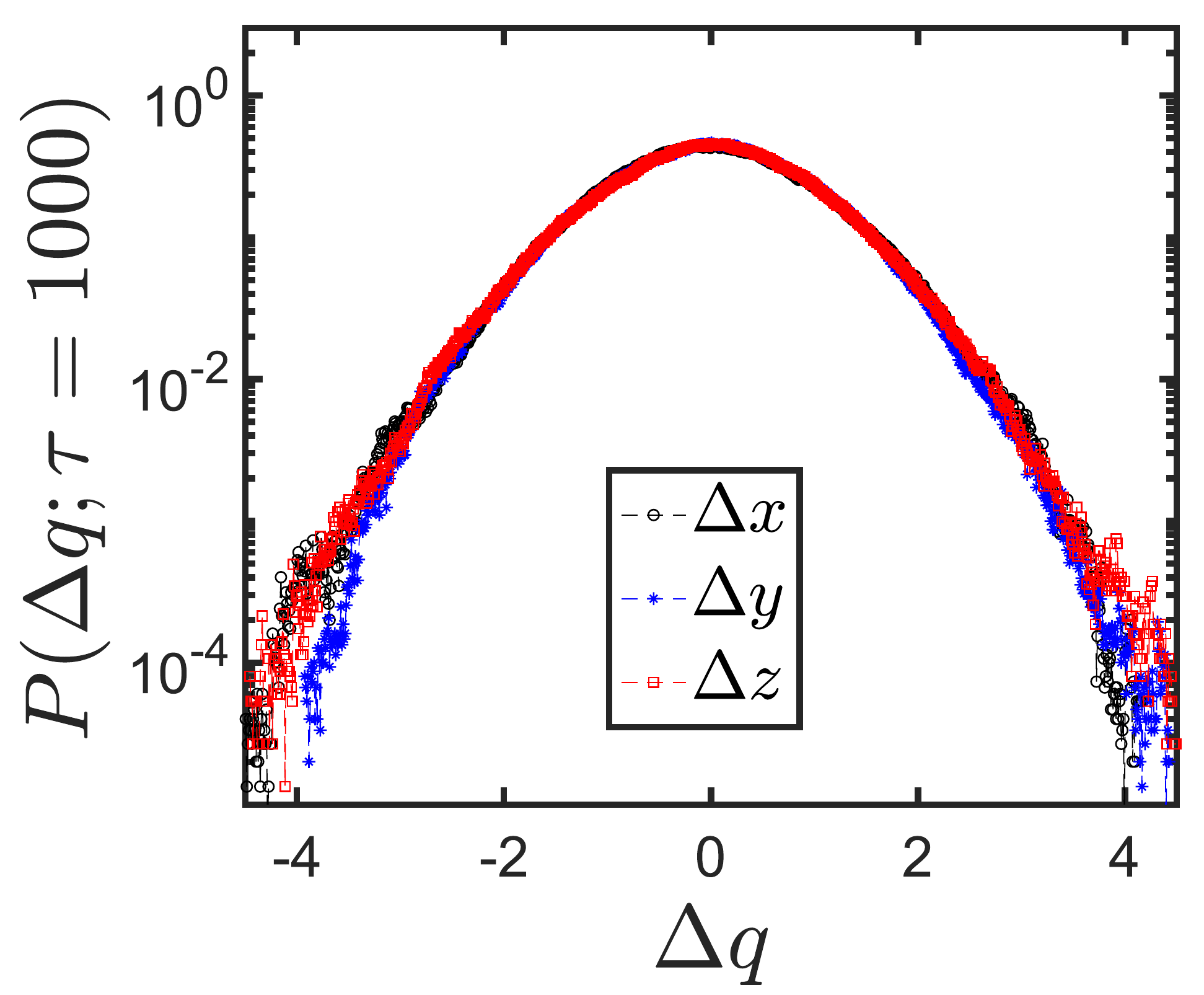} \\

                    (a) & (b) & (c) \\		
		
	\end{tabular}
\caption{Plots of the self part of the van-Hove functions in $q$, $P(\Delta q; \tau)$, where $q$ represents either of $x$, $y$ $z$, for three chosen lag-times $\tau$ (10, 100, 1000). The parameters used are $\sigma=0.5$, $\epsilon=2$ and $k=5$.}	
\label{fig:vanhoveq}
\end{figure}

\noindent To gain a deeper understanding of the underlying complex movement of the tracer probe particle in the polymer network, we analyze the probability distribution function $P(\Delta x;\tau) \equiv \left\langle \delta(\Delta x - (x (t+\tau)-x (t)))\right\rangle $ in one dimension where $x (t+\tau)$ and $x (t)$ are the positions of the tracer along $x$ direction at time $(t+\tau)$ and $t$ respectively. $P(\Delta x;\tau)$ corresponds the time-and-ensemble averaged self part of the van-Hove function. Thus the van-Hove function or the distribution is calculated from a single trajectory and that single trajectory is constructed from different individual realizations (independent trajectories). Plots of $P(\Delta x;\tau)$ are depicted in Fig.(\ref{fig:vanhove}) with the corresponding Gaussian distribution functions for the free Brownian motion, $P (\Delta x)=\frac{1}{\sqrt{2\pi\left<\Delta x^2\right>}}\exp\left({-\frac{\Delta x^2}{2\left<\Delta x^2\right>}}\right)$. For relatively smaller values of $\epsilon, \sigma$ and $k$, $P(\Delta x;\tau)$ is almost Gaussian at relatively short and long lag times and for the intermediate times it is non-Gaussian. For relatively higher values of $\epsilon, \sigma$ and $k$, $P(\Delta x;\tau)$ deviates from Gaussianity even at longer time lags which describes the trapped motion of the probe. However, due to confined motion, the distribution functions become narrower with increasing $\epsilon$, $\sigma$ and $k$. A careful observation indicates that this is not the case for $\sigma=1.5$, when the probe size is comparable to the mesh size ($\sigma_\textrm{{mesh}}\approx 1.12$) of the network. In case of this large probe $(\sigma=1.5)$, $P(\Delta x;\tau)$ deviates more from the Gaussian distribution with increasing time differences (red curve in Fig.(\ref{fig:vanhove}) (f)). This happens because for higher $\sigma$, the probe stretches the network which gives rise of the fat tails in the distribution function at longer time differences \cite{goodrich2018enhanced}. These tails go beyond the tails of the distributions for $\sigma=1$ and $\sigma=0.5$ having much higher values at that length scale ($\sigma\approx 4$) and even overlap with the distribution for the smaller tracer ($\sigma=0.3$), which is capable of moving larger distances owing to its smaller diameter without being much interrupted by the presence of polymer network. However, these are rare events because the probability associated with these large displacements are low. On the other hand due to the topological constraints for the network, though the bigger probe stretches the network but cannot escape (as in our simulations, the network cannot break). If network breakage were allowed, the bigger probe would have escaped and shown enhanced diffusion. This could be a plausible mode of transport of proteins through NPC \cite{chakrabarti2014diffusion, goodrich2018enhanced}. This is further confirmed from the results shown in Fig.(\ref{fig:vanhovediffk}) where we plot $P(\Delta x;\tau)$ for $\sigma=1.5$ at longer time differences $(\tau=100, 1000, 10000)$ with different $k (=5, 15)$ keeping $\epsilon=2$. For $k = 5$, the distribution becomes broader at longer time differences, contrary to the situation where the distribution is narrower for $k=15$ at the same time differences, when the network is too rigid even for the bigger probe to stretch efficiently. But for small probes, on increasing the $k$ which is qualitatively similar to a situation where the polymers dries up, as in polymer thin films at lower humidity with small free volume, the distributions for the small molecule probes become narrower \cite{bhattacharya2013plasticization}. This can clearly be seen from Fig.(\ref{fig:vanhove}) (g), (h) and (i).
\\
\\
We plot the distribution functions for $q$ in Fig.(\ref{fig:vanhoveq}), where $q$ represents all the three $x$, $y$ and $z$ directions. The trends are same in all the directions which means there is no preferred direction of motion of the probe contrary to the directed motion as observed in living systems \cite{chaki2019effects, du2019study}. 

\subsection{Angular distribution function}
\begin{figure}[h]
	\centering
	\begin{tabular}{ccc}
		\includegraphics[width=0.33\textwidth]{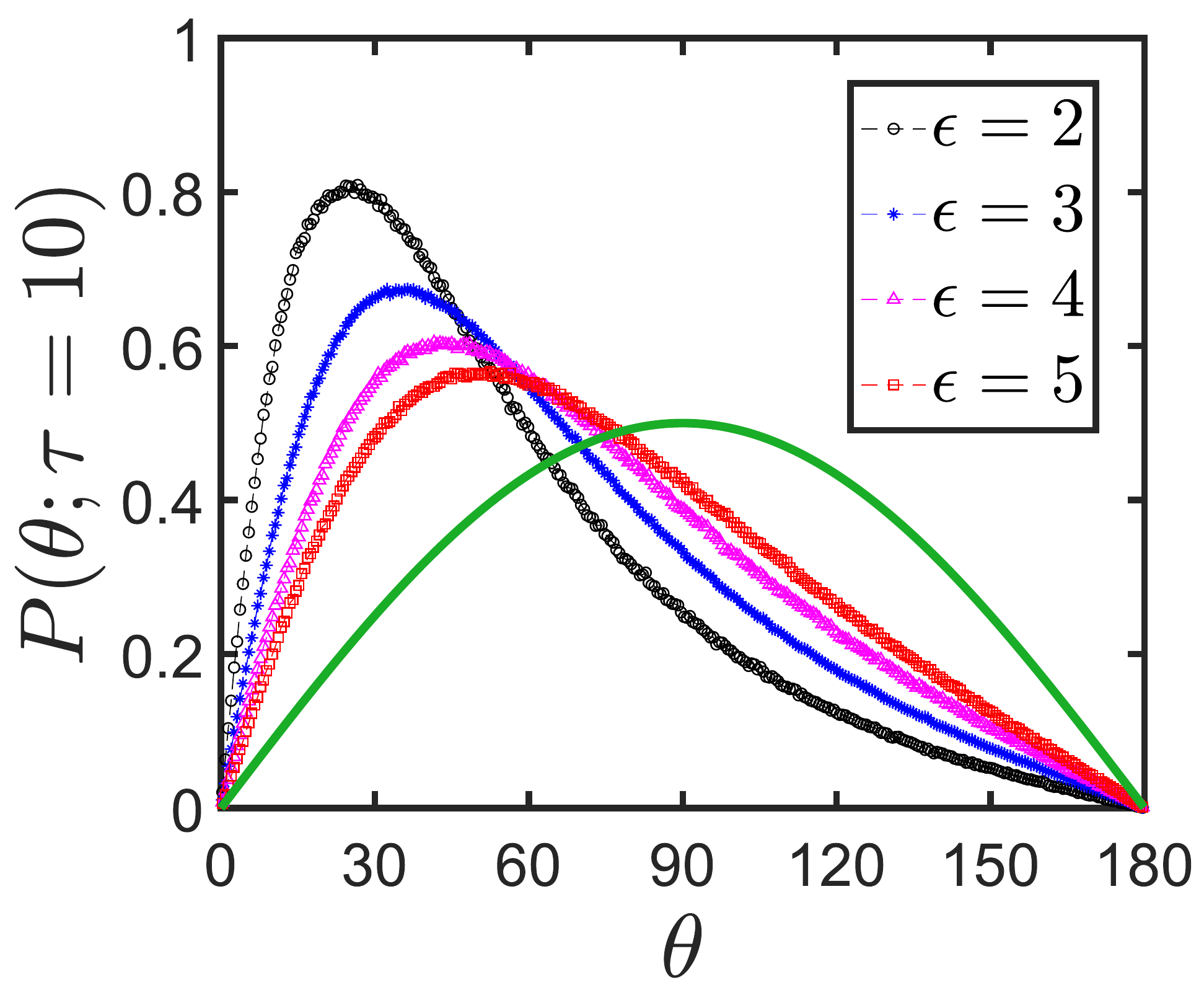} &
		\includegraphics[width=0.33\textwidth]{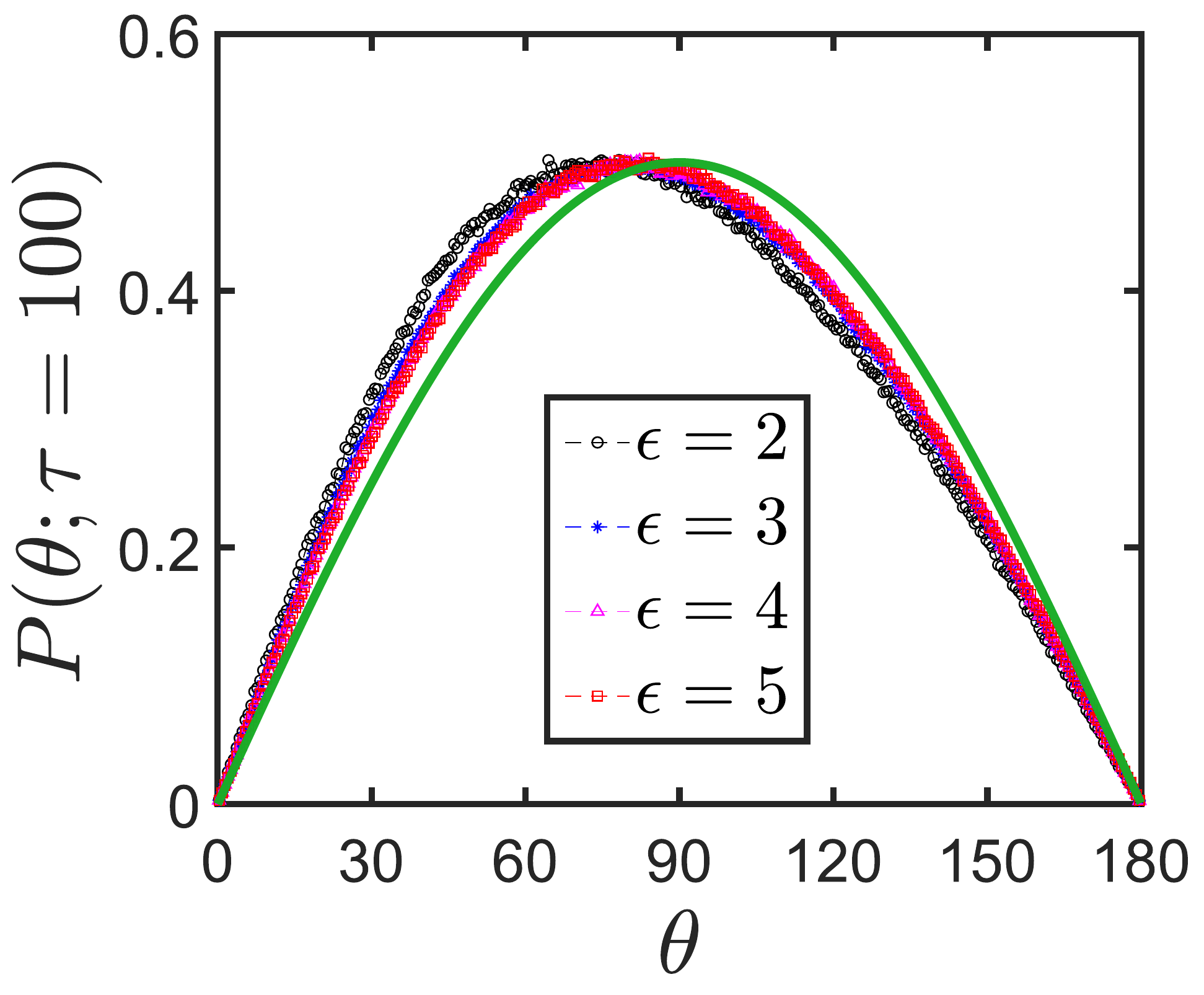} &
		\includegraphics[width=0.33\textwidth]{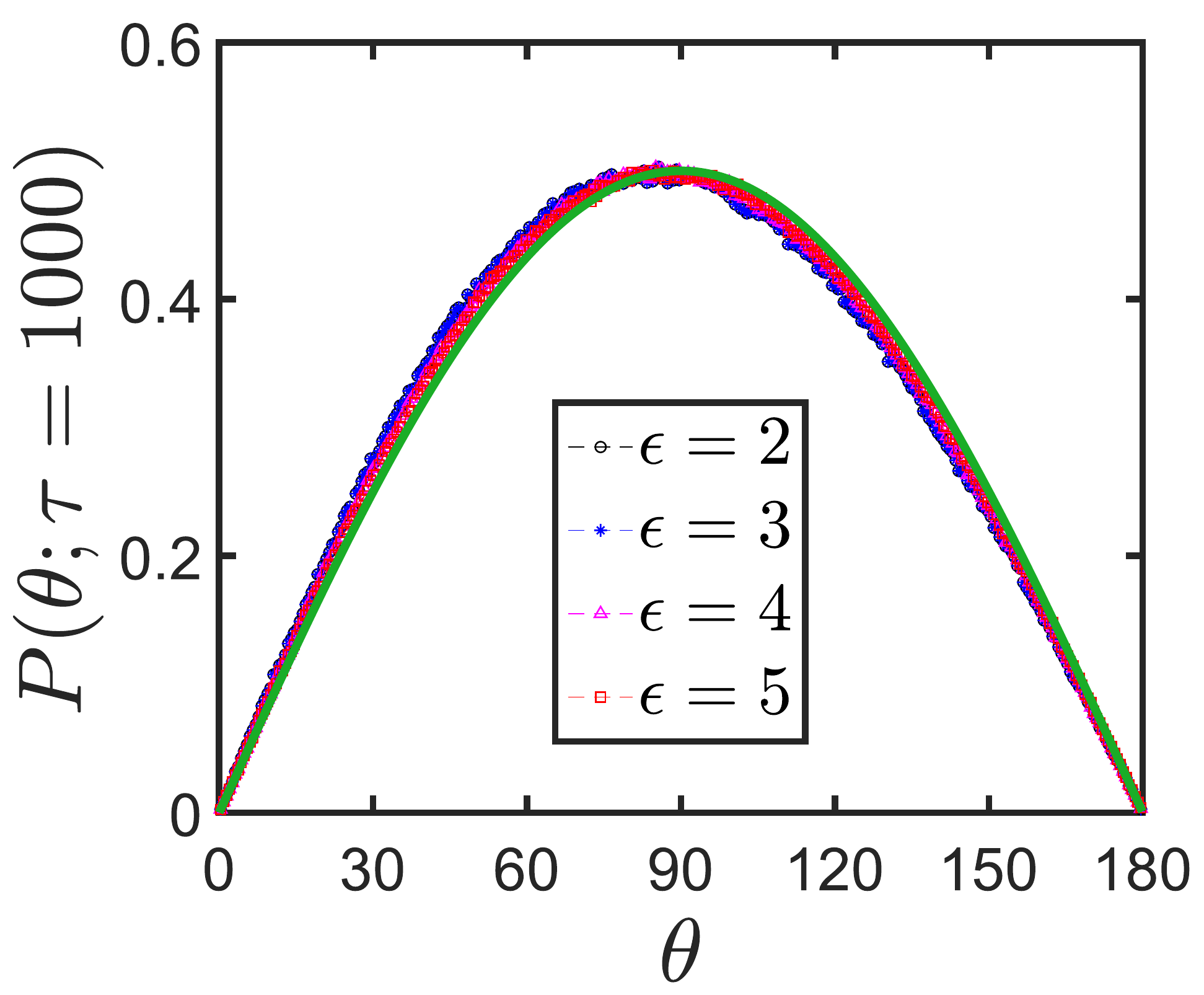}  \\
		
                  (a) & (b) & (c)  \\		
		
		\includegraphics[width=0.33\textwidth]{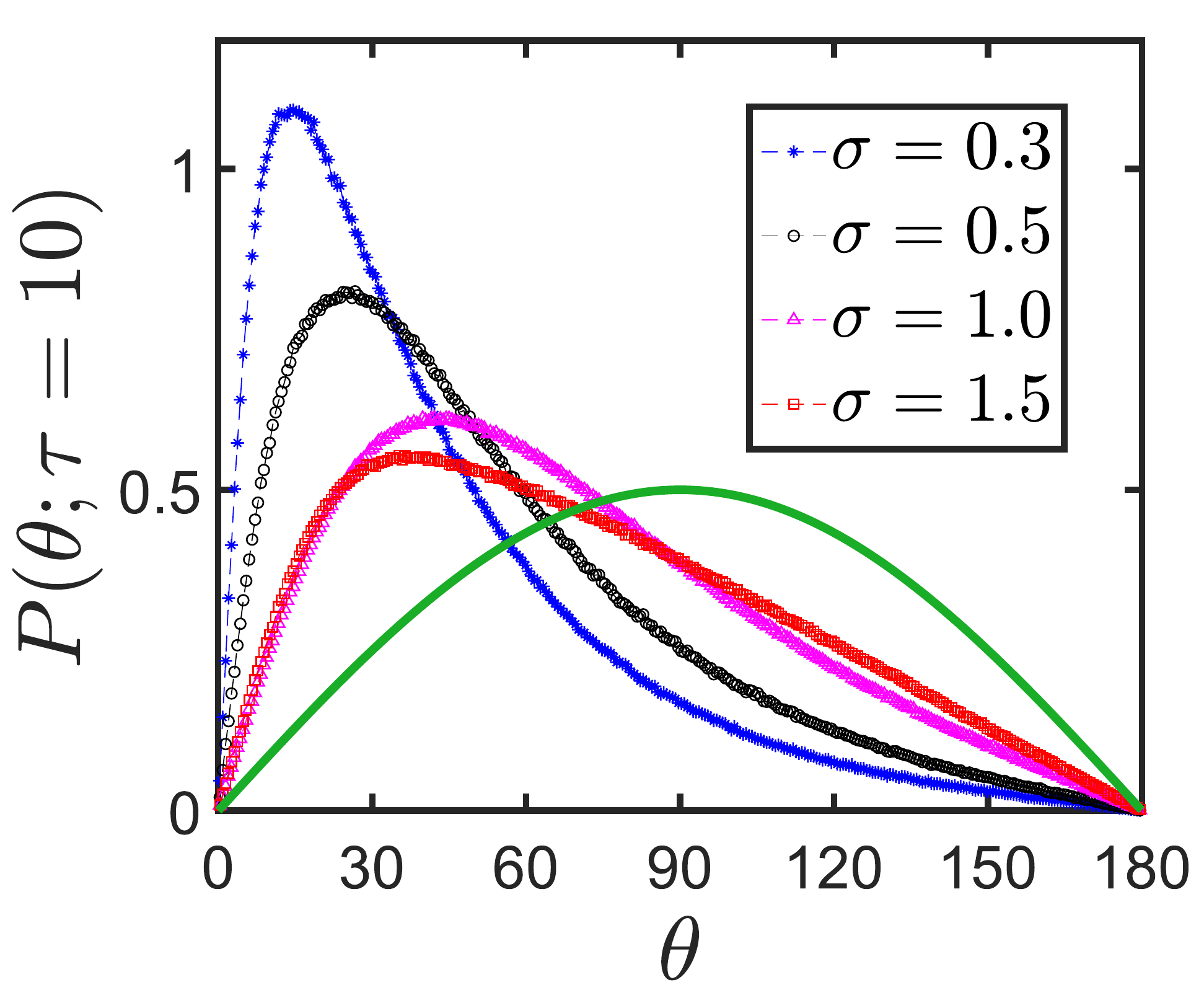} &
		\includegraphics[width=0.33\textwidth]{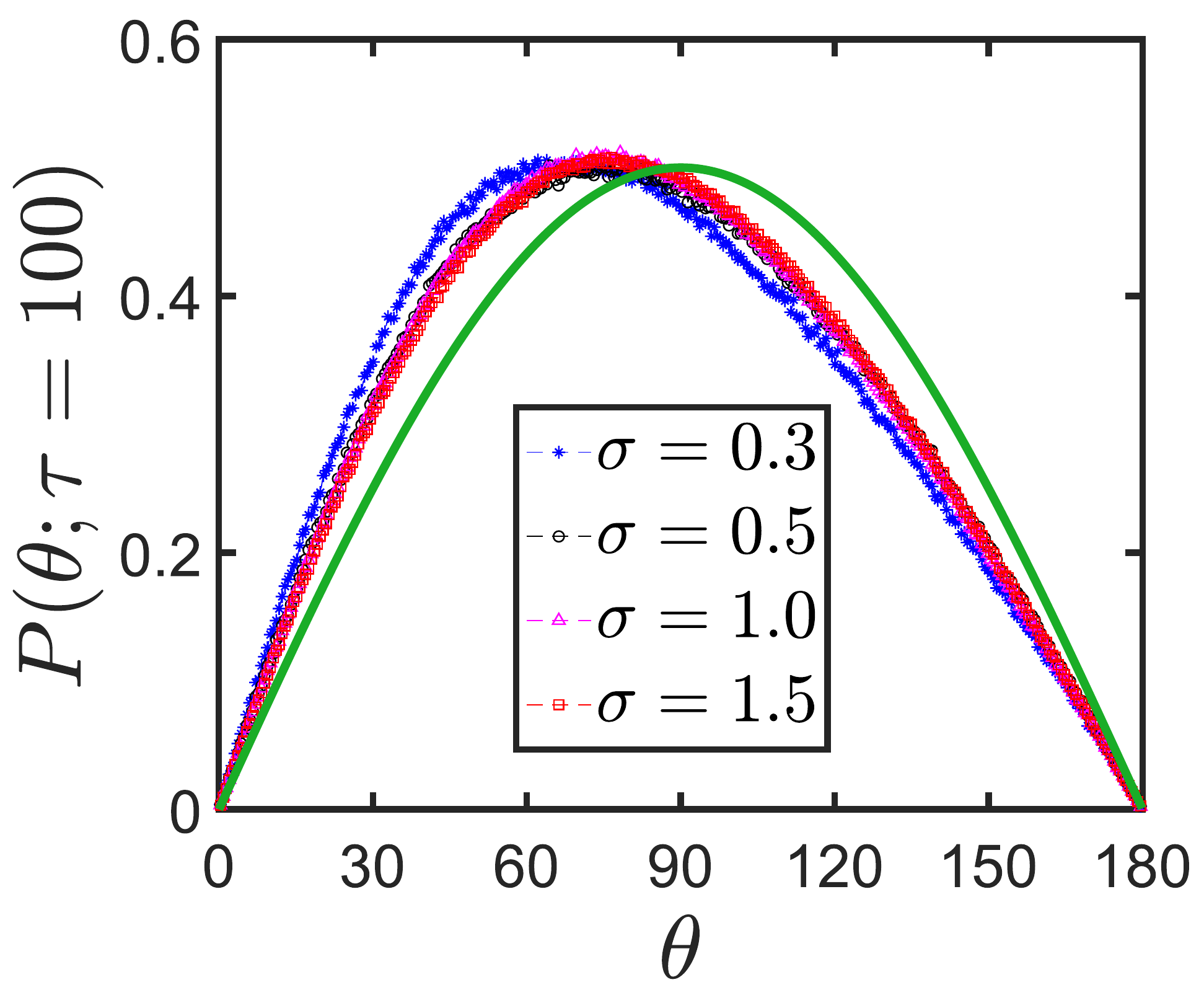} & 
		\includegraphics[width=0.33\textwidth]{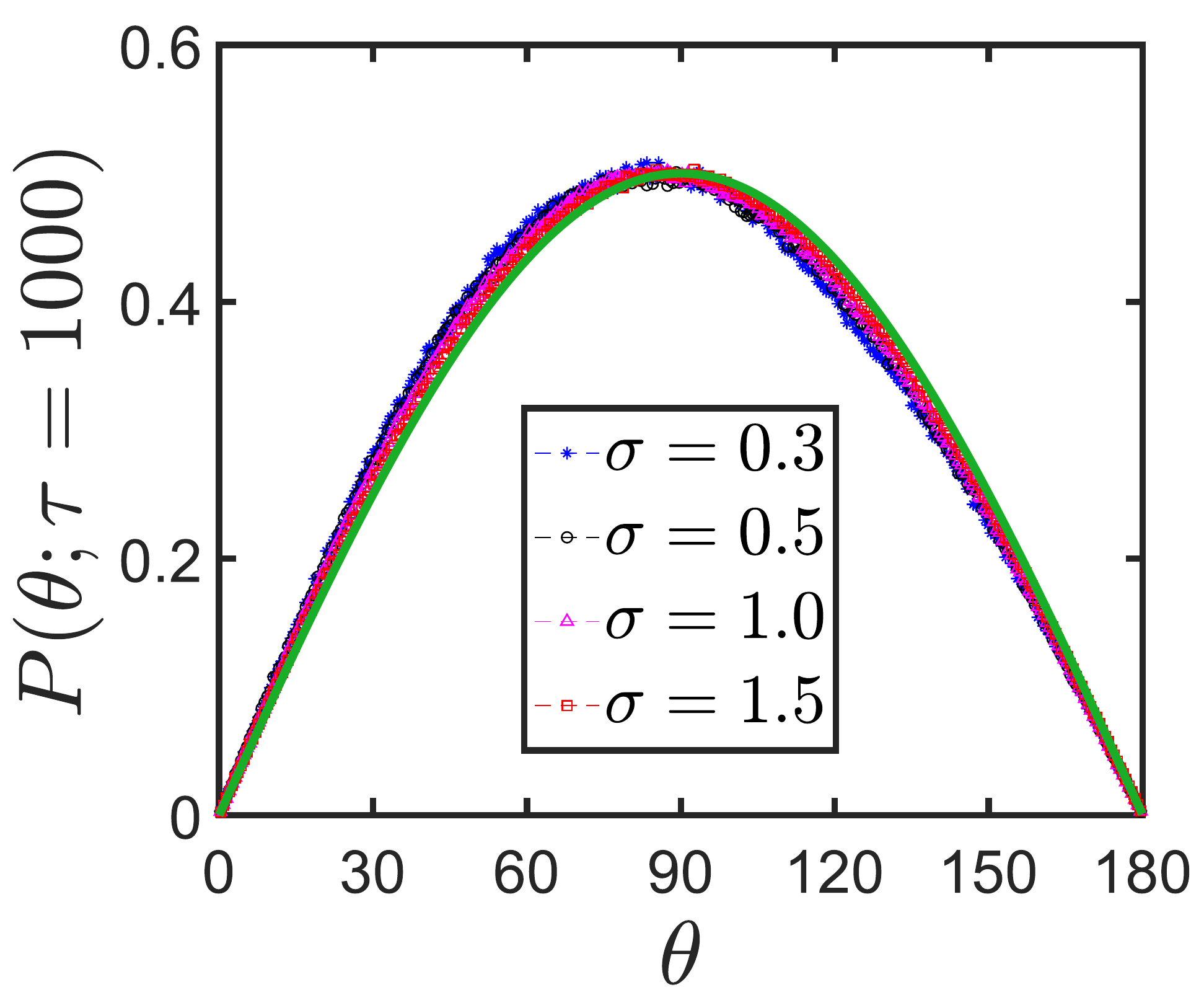}  \\
		
                  (d) & (e) & (f)  \\		
		
		\includegraphics[width=0.33\textwidth]{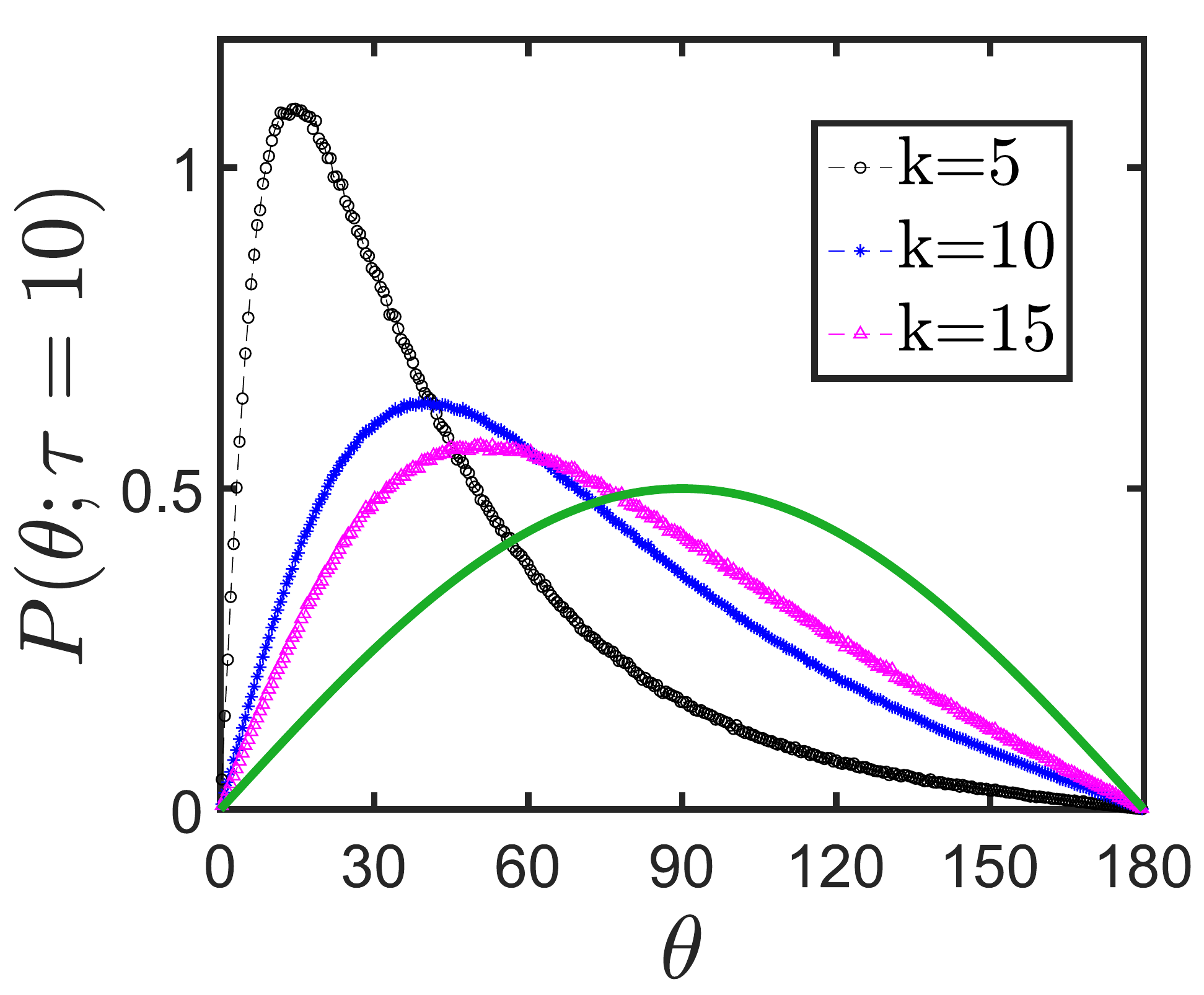} &
		\includegraphics[width=0.33\textwidth]{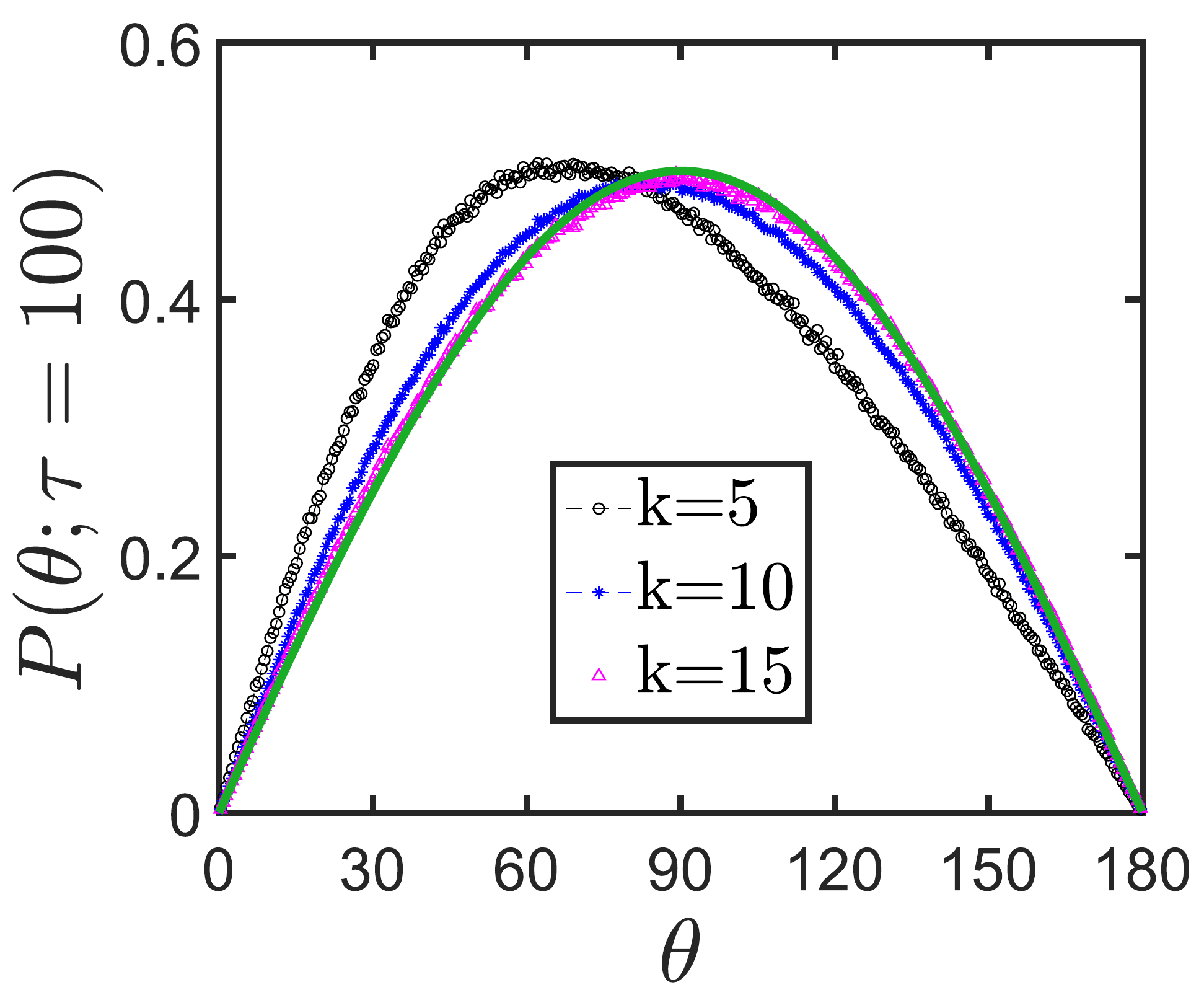} & 
		\includegraphics[width=0.33\textwidth]{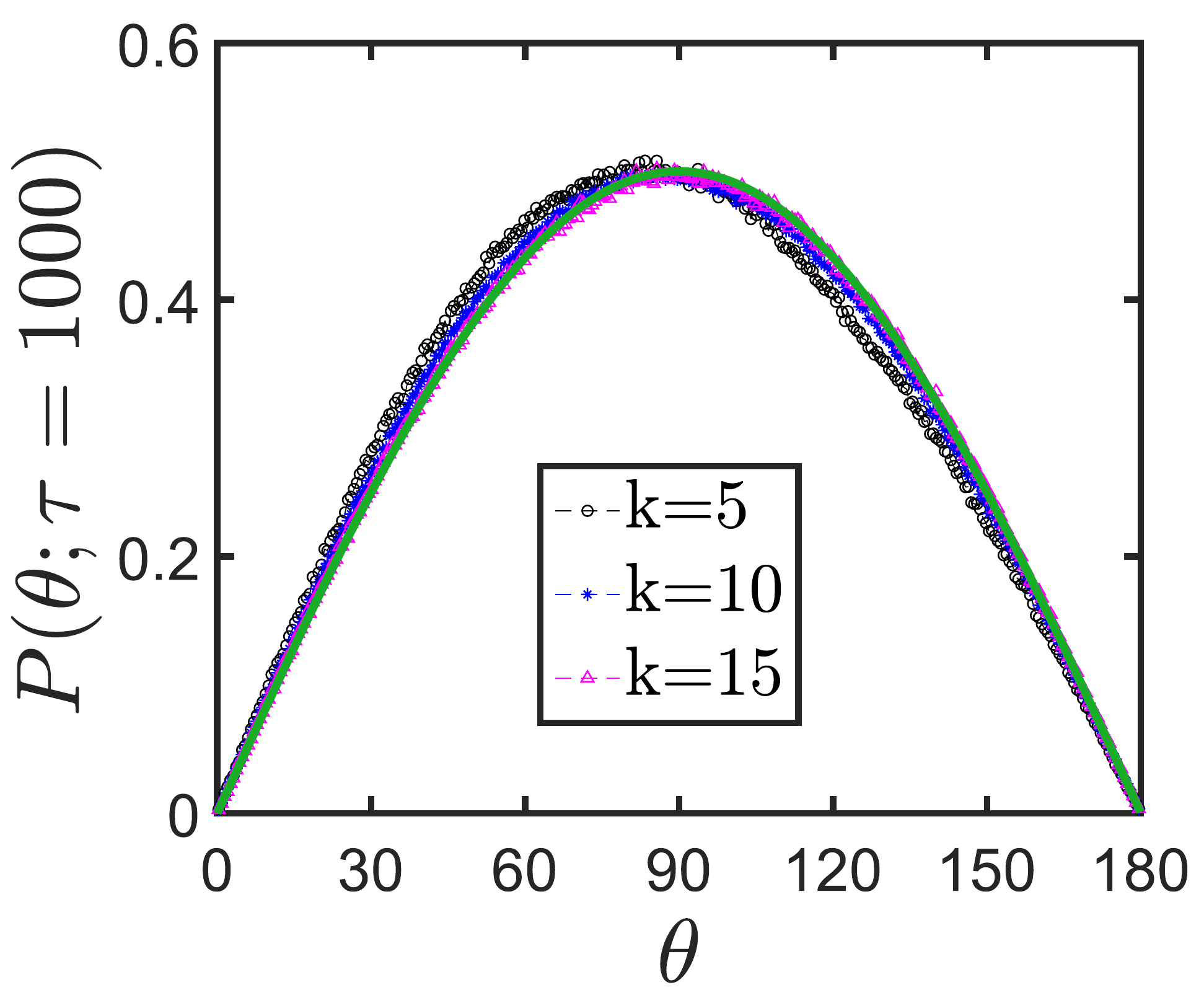} \\
		
                  (g) & (h) & (i)  \\		
		
	\end{tabular}
\caption{Angular distribution functions $P(\theta; \tau)$  for the angle $\theta$ at three lag-times $\tau$ (10, 100, 1000) for different tracer-monomer binding affinities $ (\epsilon)$ ((a), (b) and (c)) with $\sigma=0.5$, $k=5$, different sizes of the tracer ($\sigma$) ((d), (e) and (f)) with $\epsilon=2$, $k=5$ and the stiffness constants ($k$) ((g), (h) and (i)) with $\sigma=0.5$, $\epsilon=2$.The green solid lines represent the corresponding isotropic distribution $\frac{\sin\theta}{2}$.} 
\label{fig:ptheta}
\end{figure}

\noindent To quantify the trapped motion of the probe further, we consider the angular distribution function $P(\theta;\tau)$ \cite{nahali2018nanoprobe}, where the angle $\theta (\tau)$ $\equiv  \cos^{-1}\left( \dfrac{\left( \textbf{r}(t+\tau) - \textbf{r}(t)\right).\left(  \textbf{r}(t+2\tau)  - \textbf{r}(t+\tau)\right)}{\vert\textbf{r}(t+\tau) - \textbf{r}(t)\vert \vert\textbf{r}(t+2\tau) - \textbf{r}(t+\tau)\vert} \right) $  between spatial displacements separated by lag-time $\tau$ taken successively along the trajectory which essentially means each trajectory of a probe can be viewed as a snapshot of polymer conformations. Thus, $P(\theta;\tau)=\left<\delta (\theta-\theta (\tau))\right>$.
\\
\\
For isotropic displacements, the probability of finding the probe  with the orientation given by the angles $\theta$ and $\phi$ is $ \int \int \frac{1}{4\pi} \sin\theta d\theta d\phi$ where the probability distribution function $P (\theta,\phi)=\frac{1}{4\pi}$ for the purely isotropic case. With this, it is straightforward to show that $P(\theta)=\frac{\sin\theta}{2}$ for isotropic displacements. Hence, any deviations from $\frac{\sin\theta}{2}$ would be a measure of correlated motion of the probe. Fig. (\ref{fig:ptheta}) shows significant deviation from $\frac{\sin\theta}{2}$ at short times reflecting the trapped or in other words correlated motion of the probe. However, at longer time differences, $P(\theta;\tau)$ almost merges to $\frac{\sin\theta}{2}$ for all the parameters showing the long time nearly isotropic diffusion of the probe.  For short time with increasing values of $\sigma, \epsilon, k$,  which facilitate trapping, $P(\theta;\tau)$ shifts more towards right which is compatible with the picture where the probe revert their motion frequently as a consequence of the viscoelastic response of the medium or the jiggling type motion inside the cages. For example in Fig. (\ref{fig:ptheta}) (a), the curve for $\epsilon=5$ (red) has the peak at a higher $\theta$ compared to any other lower $\epsilon$. However the plots for the same set of parameter at higher $\tau=1000$ almost overlap with each other and also with the $\frac{\sin\theta}{2}$  curve as can be seen from Fig. (\ref{fig:ptheta}) (c). This confirms that though the probe motion has some short time heterogeneity, in the long time it is homogeneous. This is consistent with the fact that the polymer network is on a lattice which is globally homogenous (beyond a length scale) but has local heterogeneity. This local heterogeneity is probed in the short time but no signature of this heterogeneity remains in the long time.

\section{Conclusions} \label{sec:conclusion}

\noindent A large number of transport processes in biology and materials science occur in crowded medium. Gel-like materials form a subclass of such crowded media, examples of which include the central plug of NPC \cite{goodrich2018enhanced, chakrabarti2014diffusion, gopinathan2017}, mucus membrane \cite{lieleg2012mucin, lai2009mucus}, polymer thin films \cite{flier2011heterogeneous, bhattacharya2013plasticization}, actin networks \cite{phillips2012physical, mizuno2007nonequilibrium, sonn2017scale}.  In this paper we analyze the effect of the size of the probe, stickiness of the probe to the network and the rigidity of the network on the probe dynamics in a polymer network (gel), constructed on a diamond lattice. The diamond lattice provides substantial crowding but ensures homogeneity, beyond a length scale which is accessible in moderate to long time. Thus our system is quite different from systems with inherent heterogeneity, where different regions have different mobilities \cite{samanta2016tracer, jain2016diffusion, chechkin2017brownian}. In general, on increasing the stickiness, probe size and the network rigidity, dynamics of the probe slows down, becomes more and more subdiffusive with narrower  non-Gaussian distributions in the short to intermediate time. In addition, in the short time, the velocity autocorrelation functions have negative dips owing to caging of the probe, where FBM and CTRW both contribute.  However, on increasing the rigidity of the network, motion of the probe becomes more confined as experimentally observed in the case of small molecule transport in polymer thin films at lower humidity \cite{bhattacharya2013plasticization} or on lowering the temperature towards the glass transition temperature ($T_g$) \cite{flier2011heterogeneous}.  On the other hand, to our surprise, for a probe comparable to the mesh size of the network with moderate stickiness, the long time displacement distribution shows fat tails confirming stretching of the network. However, these are rare events since the associated probabilities are quite low due to the topological constraint on the network. In other words, the network can stretch to a finite length but cannot break. This could be an important mode of transport for larger probes in a gel-like environment in general. 
\\
\\
\noindent We hope that our study, based on molecular dynamics simulation of probes in a polymer network will shed some light in understanding a large number of phenomena involving transport of probe particles (from molecular to nano sized) through gel like medium and in crowded medium in general. 

\section{Acknowledgements}

\noindent RC acknowledges SERB for funding (Project No. SB/SI/PC-55/2013). PK, LT thank UGC for fellowships. SC thanks DST-Inspire for a fellowship. Authors thank R. Kailasham for helping with plotting figures and critically reading the manuscript.

%\begin{bibliography}
\bibliographystyle{apsrev}

%\usepackage[square,numbers,sort&compress]{natbib}
%\bibliographystyle{apsrev}
%\usepackage{doi}
%\usepackage{hyperref}
%\bibliography{gel_transport}
%\end{bibliography}
\end{document}